\documentclass[referee]{raa}

\usepackage{graphicx,times}
\usepackage{natbib}
\usepackage{amssymb,amsmath}
\bibpunct{(}{)}{;}{a}{}{,}
\usepackage[pagebackref=true]{hyperref}

\newcommand{\HII}{H\,{\sc ii}}

\newcommand{\Ha}{H$\alpha$\,$\lambda$6563}
\newcommand{\Hb}{H$\beta$\,$\lambda$4861}

\newcommand{\OIII}{[O\,{\sc iii}]}

\newcommand{\LOIII}{$L_{\rm [O\,{\textsc{iii}}]}$}
\newcommand{\OIIII}{{\rm [O\texorpdfstring{\,}{}{\sc iii}]\texorpdfstring{\,}{}\texorpdfstring{$\lambda$}{}5007}}
\newcommand{\NII}{[N\,{\sc ii}]\,$\lambda$6583}

\begin{document}

   \title{Stellar Populations of AGN-host Dwarf Galaxies Selected with Different Methods}

 \volnopage{ {\bf 20XX} Vol.\ {\bf X} No. {\bf XX}, 000--000}
   \setcounter{page}{1}
   \author{Xiejin Li\inst{1,2}, 
           Yinghe Zhao\inst{1,3,4}\footnote{Corresponding author},
           \and
           Jinming Bai\inst{1,4}
           }

   \institute{Yunnan Observatories, Chinese Academy of Sciences, Kunming 650216, China; {\it zhaoyinghe@ynao.ac.cn}\\
        \and
            University of Chinese Academy of Sciences, Beijing 100049, China\\
        \and
        Key Laboratory of Radio Astronomy and Technology (Chinese Academy of Sciences), A20 Datun Road, Chaoyang District, Beijing, 100101, P. R. China\\
        \and
        Key Laboratory for the Structure and Evolution of Celestial Objects, Chinese Academy of Sciences, Kunming 650011, China\\
\vs \no
   {\small Received 20XX Month Day; accepted 20XX Month Day}
            }

\abstract{In this paper we investigate the stellar populations and star formation histories of 235 active galactic nuclei (AGN)-host dwarf galaxies, consisting of four samples identified separately with different methods (i.e., radio, X-ray, mid-IR and variability), utilizing the synthesis code STARLIGHT and spectra from the Sloan Digital Sky Survey (SDSS) Data Release 8. Our results show that the variability sample is the oldest, while the mid-IR sample is the youngest, for which the luminosity at 4020 \AA\ is dominated ($>50\%$) by the young population ($t<10^8$ yr). The light-weighted mean stellar age of the whole sample is in general about 0.7 dex younger than the optical sample studied in Cai et al. We compare the population results between fitting models with and without a power-law (PL) component and find that the neglect of a PL component would lead to an under- and over-estimation by 0.2 and 0.1 dex for the light- and mass-weighted mean stellar age, respectively, for our sample of dwarf galaxies, which has a mean fractional contribution of $\sim$16\% from the AGN. In addition, we obtain further evidence for a possible suppression of star formation in the host galaxy by the central AGN. We also find that there exists an anti-correlation between the extinction-corrected \OIII\ luminosity and light-weighted mean stellar age, confirming our previous finding that there is a physical connection between AGN and star-forming activities in AGN-host dwarfs.
}
   \authorrunning{Xiejin Li et al. }  
   \titlerunning{Stellar Populations of AGN-host Dwarf Galaxies}  
   \maketitle

\section{Introduction}  
\label{sect:intro}
Dwarf galaxies, which constitute the largest population of galaxies in the present-day universe \citep{Marzke1997}, are characterized by diverse morphologies and star formation histories (SFHs;  \citealt{Tolstoy2009,McQuinn2011,weisz2011a,Gallart2015}). According to the hierarchical clustering model, dwarf galaxies serve as the fundamental building blocks of massive galaxies in the universe \citep{Kauffmann1993}. Black holes (BHs) in low-mass dwarf galaxies may be similar to the first seed BHs \citep{Bellovary2011} owing to their relatively quiescent merger histories. Furthermore, dwarf galaxies are generally metal-poor \citep{Mateo1998}, making them similar to high-redshift objects. Therefore, dwarf galaxies are important objects for studying the formation and evolution of galaxies, as well as in cosmological simulations.

The evolution of dwarf galaxies may be influenced by both external (such as mergers and interactions) and internal (e.g., active galactic nucleus; AGN) factors. For massive galaxies, AGNs play a crucial role in star formation and evolution through different feedbacks on the host galaxies, which has been studied comprehensively by observational works and numerical simulations. Some studies suggest that AGNs are likely to disperse gas within galaxies, thereby inhibiting star formation \citep{Vogelsberger2014,Schaye2015,Dubois2013,Dubois2016,Pillepich2018}. In contrast, some other works show that AGNs may also compress gas, hence triggering star formation \citep{Silk2009,Gaibler2012,Ishibashi2012,Nayakshin2012,Kalfountzou2012,Silk2013,Kalfountzou2014,Querejeta2016,Maiolino2017,2022Natur.601..329S}. To explore the possible connection between AGNs and host galaxies, therefore, it is important to study the stellar populations and star formation histories of AGN-host galaxies (e.g., \citealt{Cid2004, 2018ApJ...864...32J, 2018MNRAS.478.5491M, Cai2020, Cai2021, 2022ApJ...926..184J}).

Abundant works in literature have systematically investigated the stellar populations of normal dwarf galaxies \citep{Hodge1989,Mateo1998,Dolphin2005,Tolstoy2009,McQuinn2010,Weisz2011b,Zhao2011,Kauffmann2014,2017ApJS..233...13Z,Zheng2017,Taibi2022}, with a significant focus on objects located within the local group (see \citealt{Tolstoy2009} for an overview). As shown in  \citet{Tolstoy2009}, early-type dwarf galaxies, such as dwarf spheroidal galaxies (dSphs), have typically not formed stars for at least several hundred million years. \citet{Weisz2011b} analyze the SFHs of 60 dwarf galaxies in the nearby Universe ($\lesssim$4 Mpc) and find that the majority of dwarf galaxies formed the bulk of their mass prior to redshift $z\sim$1. The authors also show that dwarfs with different morphological types exhibit similar SFHs when age less than 1 Gyr, while stellar mass formed in recent 1 Gyr is correlated with morphology. A series of works (e.g., \citealt{Harbeck2001,Tolstoy2004,Battaglia2006,Bernard2008,Belfiore2018,Taibi2022}) have also identified a radial gradient in the stellar age and/or metallicity of dwarf galaxies.

Compared with normal low-mass galaxies, however, there are only few studies devoted to investigating the stellar populations of AGN-host dwarfs. \citet[hereafter C20]{Cai2020} analyze the stellar populations and SFHs for 136 AGN-host dwarf galaxies from \citet[hereafter R13]{Reines2013} selected by the optical emission-line ratio diagnosis diagram\citep[i.e., the BPT diagram;][]{Baldwin1981}, and find that these optically selected AGN-host dwarfs show a wild diversity of SFHs and may have repeated star-forming activities. The authors also obtain a mild correlation between SFH and \OIIII\ line luminosity (\LOIII) for objects with $L_{\mathrm{[O\,{\textsc{iii}}]}} > 10^{39}$~ erg~s$^{-1}$, indicating a physical connection between AGN and star-forming activities. However, the old population ($t>10^9$ yr) contributes most light for the majority of their sample galaxies and dominates the stellar mass, which is supported by the analysis of the radial stellar populations for a sample of 60 optically selected AGN-host dwarf galaxies \citep{Cai2021}. 

However, AGNs selected by the BPT diagram generally show a redder $g-r$ color \citep[R13;][C20]{Moran2014,Sartori2015}, indicating a selection effect likely caused by the fact that the optical diagnostics is not sensitive to AGNs when the star-forming host galaxies dominates the emission-line spectra (e.g., \citealt{Trump2015,2019ApJ...870L...2C}). 
In other words, the light contribution from AGNs can be heavily contaminated by the host galaxy with ongoing star formation, making it difficult to identify the internal AGN except for red, quenched systems, in which the features of BH accretion can be more prominent. This is more serious for dwarf galaxies, which generally have an lower metallicity \citep{Groves2006}, leading to
a lower \NII/\Ha\ ratio and a smaller \OIIII/\Hb\ spread \citep{Ludwig2012,Kewley2013}, and thus are more likely to overlap with star-forming galaxies on the top-left of the BPT diagram. As a result, the suppression of star formation by AGNs could be overestimated in AGN-host dwarf galaxies. 

In order to reduce/overcome the influence of selection bias suffered by the optical emission-line method, and to further establish connections between AGNs and hosts, it is necessary to systematically investigate the stellar populations and SFHs of AGN-host dwarfs selected through different techniques. To this end, in this paper we present a detailed study of the stellar populations for a sample of 235 AGN-host dwarf galaxies,  consisting of four subsamples selected with different methods/bands (namely radio, mid-infrared, variability and X-ray), using a simple stellar population (SSP) synthesis method that is capable of yielding the various stellar components, AGN contribution to the optical continuum, and internal extinction. The paper is organized as follows: Section \ref{sec:Sample and Data Reduction} describes the selection of the AGN-host dwarf galaxies, properties of different samples and data reductions. Stellar population synthesis results and discussion are given in Section \ref{sec:Result and Discussion}. We summarize our results in Section \ref{sec:Summary}. Where required we adopt a Hubble constant of $H_0 = 73$~km~s$^{-1}$~Mpc$^{-1}$.

\section{Sample and Data Reduction} \label{sec:Sample and Data Reduction}
\subsection{Sample Selection} \label{subsec:sample select}

Here we compile a sample of AGN-host dwarf galaxies from four works that select AGNs using the four aforementioned methods. We briefly describe the sample selection for each subsample in the following. 

{\it The Radio Sample (hereafter radio):} \citet{Reines2020} observed 111 galaxies  ($M_\star \lesssim 3 \times 10^9~M_\odot$ and redshift $z<0.055$), selected by cross-matching the NASA-Sloan Atlas (NSA v0\_1\_2) with the Faint Images of the radio Sky at Twenty centimeters (FIRST) Survey catalogue (version 2013 Jun 5), using sensitive, high-resolution observations from the Karl G. Jansky Very Large Array (VLA). After carefully evaluated possible origins for the radio emission, including thermal \HII\ regions, supernova remnants, younger radio supernovae, background interlopers, and AGNs, in the target galaxies, they identify 13 out of 39 compact radio sources that almost certainly host active massive BHs. We adopt these 13 galaxies as the radio-selected sample.

{\it The X-ray Sample (hereafter X-ray):} \citet{Birchall2020} present a sample of 61 dwarf galaxies ($M_\star \lesssim 3 \times 10^9 M_\odot$ and $z<0.25$) that exhibit nuclear X-ray activity indicative of an AGN. This sample of galaxies are identified from a parent sample of 4331 dwarf galaxies find within the footprint of both the MPA-JHU catalogue (based on SDSS DR8) and 3XMM DR7, after applying the following criteria: (1) the extent of the X-ray source is less than 10\arcsec\ to ensure point-like emission that is consistent with an AGN \citep{Rosen2016}; and (2) there is an X-ray excess relative to the contribution, estimated with the star formation rate (SFR), from X-ray binaries \citep{Lehmer2016} and hot gas in the interstellar medium \citep{Mineo2012}. These 61 X-ray-selected dwarf galaxies constitute our X-ray (parent) sample.

{\it The Mid-Infrared (IR) Sample (hereafter mid-IR):} In \citet{Sartori2015}, the authors identify AGN candidates using mid-IR color cuts (\citealt{Stern2012,Jarrett2011}). A source is classified as an AGN when it meets: (1) $[W1-W2]\geqslant0.8$; and/or (2) $2.2<[W2-W3]<4.2\ \&\ (0.1\times[W2-W3]+0.38)<[W1-W2]<1.7$ (\citealt{Sartori2015}). Based on a parent sample of 48416 dwarf galaxies ($M_\star \lesssim 10^{9.5} M_\odot$ and $z<0.1$), obtained by cross-matching the OSSY catalogue \citep[Oh–Sarzi–Schawinski–Yi;][]{Oh2011} with the MPA-JHU catalogue, 189 AGNs are identified using the data observed by the Wide-field Infrared Survey Explorer (WISE) Telescope.

{\it The Variability-selected Sample (hereafter variability):} This sample consists of 192 variability-selected AGNs from \citet{Ward2022}, where the dwarf galaxies have $M_\star < 10^{9.75} M_\odot$ and $0.02<z<0.35$ in the NSA v1\_0\_1 catalogue. The authors cross-matched these dwarfs with two parent samples: the Zwicky Transient Facility (ZTF) optical catalogue, and the WISE mid-IR catalogue. They find 44 out of 25714 dwarf galaxies have optical variability, and 148 (with 2 sources overlapping with the optically variable sample) out of 79879 dwarf galaxies show mid-IR variability, after performing a statistically significant variability test and removing poor quality photometry and possible supernova objects.

In total, we obtain a parent AGN sample of 437 objects, consisting of 13 radio-, 55 X-ray-, 179 mid-IR-, and 190 variability-selected AGNs, after cross-matching all of the above AGN samples with the NASA-Sloan Atlas\footnote{\url{https://www.sdss.org/dr16/manga/manga-target-selection/nsa}} (NSA) version v1\_0\_1 catalogue, which re-analyzed the SDSS imaging and spectroscopic data using techniques in \citet{Blanton2011} and \citet{Yan2011}. These 437 sources were further applied a stellar mass (``ELPETRO\_MASS" in the NSA catalogue) cut of $M_\star < 10^{9.5}~M_\odot$ and a redshift cut of $z<0.055$ to keep consistent with the optical sample of R13 and cross-matched with the SDSS spectroscopic catalogue. Finally, a total sample of 235 unique AGN-host dwarf galaxies (hereafter the whole sample) is obtained after removing a few objects having spectra masked with bad flags, among which twelve are radio-selected from \citet[hereafter R20]{Reines2020}, forty are X-ray-selected from \citet[hereafter B20]{Birchall2020}, ninety-seven are mid-IR-selected  from \citet[hereafter S15]{Sartori2015} (two also in the radio-selected sample, and two in the variability-selected sample) and ninety are variability-selected from \citet[hereafter W22]{Ward2022}. We also note that six (three mid-IR-selected and three X-ray-selected) sources are also included in the R13 optical sample.

\begin{table}
\bc
\begin{minipage}[]{140mm}
\caption[]{Basic Information of Different Samples\label{tab1}}\end{minipage}
\setlength{\tabcolsep}{4pt}
\small
 \begin{tabular}{cccccc}
  \hline\noalign{\smallskip}
Sample & Method & Number & log($M_\star / M_\odot$) & $z$ & $g-r$ (mag)\\
  \hline\noalign{\smallskip}
R20 & Radio & 12 (13) & 9.22$\pm$0.20 (9.18$\pm$0.28) & 0.029$\pm$0.007 (0.027$\pm$0.008) & 0.35$\pm$0.11 (0.34$\pm$0.12)\\
B20 & X-ray & 40 (55) & 9.21$\pm$0.28 (9.31$\pm$0.32) & 0.028$\pm$0.010 (0.029$\pm$0.012) & 0.31$\pm$0.10 (0.32$\pm$0.12)\\
S15 & mid-IR & 97 (179) & 8.70$\pm$0.50 (8.90$\pm$0.47) & 0.029$\pm$0.014 (0.046$\pm$0.034) & 0.26$\pm$0.19 (0.18$\pm$0.27)\\
W22 & Variability & 90 (190) & 9.17$\pm$0.41 (9.47$\pm$0.28) & 0.023$\pm$0.017 (0.032$\pm$0.023) & 0.36$\pm$0.09 (0.38$\pm$0.12)\\
R13 & Optical & 136 & 9.34$\pm$0.21 & 0.028$\pm$0.011 & 0.53$\pm$0.13\\
All &   & 235 (437) & 8.95$\pm$0.53 (9.22$\pm$0.50) & 0.026$\pm$0.014 (0.034$\pm$0.023) & 0.31$\pm$0.14 (0.32$\pm$0.16)\\
  \noalign{\smallskip}\hline
\end{tabular}
\ec
\tablecomments{0.86\textwidth}{Data in parenthesis show the results of parent samples. Columns 4-6 list the median values of stellar mass, redshift and $g-r$ color with the 1 $\sigma$ dispersion estimated using $1.48\times$MAD where MAD is the median absolute deviation of the sample, and the dispersion of a mean value is estimated using the standard deviation of the sample throughout the paper.}
\end{table}

\subsection{Sample Properties} \label{subsec:Sample Properties}
\begin{figure}[tb]
   \centering
   \includegraphics[width=0.95\textwidth, angle=0]{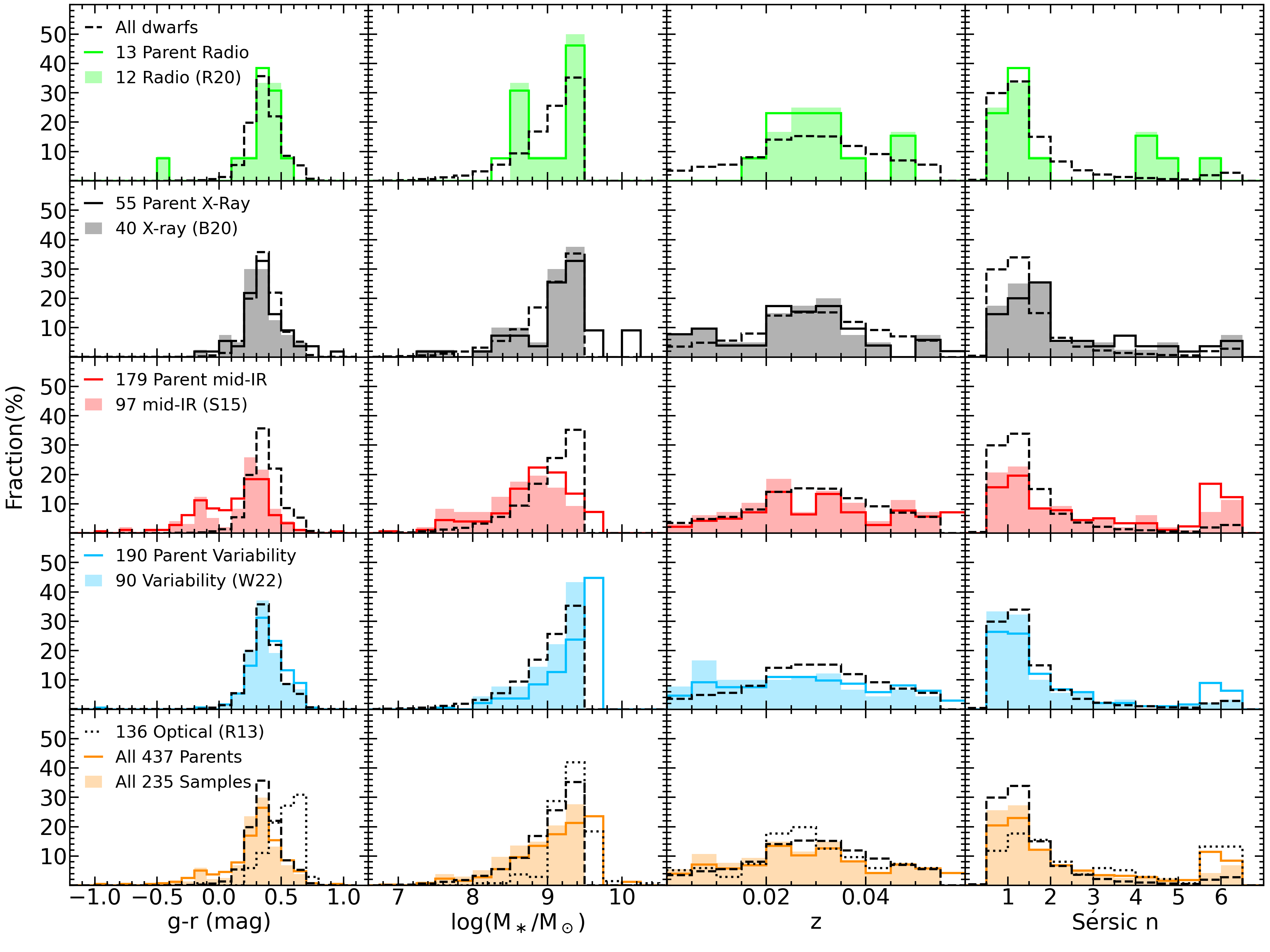}
   \caption{Rows 1-5 show the parameters distributions for distinct AGN-host dwarf galaxy samples (colored filled histograms) and their parent samples (colored solid lines) from top to bottom are radio (lime), X-ray (gray), mid-IR (red), variability (light blue) and all 235 (437) AGN-host dwarf (parent) galaxies (orange). The black dashed line shows the distribution of all dwarfs consisting of 33728 galaxies with stellar mass $M_\star < 10^{9.5} M_\odot$ and redshifts $0.001 \leqslant z<0.055$ from NSA v1\_0\_1 catalogue. The optical sample (black dotted line) is plotted in row 5 (bottom row) for comparison. Each column from left to right shows the $g-r$ color, stellar mass, redshift and S\'{e}rsic index. The $g-r$ color has been corrected for the Galactic extinction.} 
   \label{fig1}
\end{figure}

Though the current sample is not complete in any sense, it is the largest sample of AGN-host dwarfs with optical spectroscopic data to date. Our sample galaxies span a large range of galactic parameters (i.e., $g-r$ color, stellar mass and morphology type represented by the S\'{e}rsic index $n$), as plotted in Fig. \ref{fig1}, which generally show a similar distribution (color-filled histogram) to the parent sample (solid-line histogram). These results indicate that our sample can be representative of the parent AGN-host dwarf sample.

Compared the total spectroscopic sample with the NSA dwarf galaxy sample (hereafter ``All dwarfs"; dashed-line histogram), they seem to have a similar distribution in the $g-r$ color, as shown in the bottom left panel in Fig. \ref{fig1} (also see Table \ref{tab1}), in contrast to the optical sample (dotted-line histogram) that has a relatively redder color. On closer inspection, a tail towards bluer $g-r$ color can be found in our whole sample, as well as that there exists a bump (more prominent for the parent AGN sample) around S\'{e}rsic index $n=6$ (bottom right panel in Fig. \ref{fig1}). These results likely suggest that there might exist some intrinsic differences between AGN-host and normal dwarfs. Given the limited size of the AGN-host dwarfs, however, a much larger sample is needed to reach a solid conclusion.

In Table \ref{tab1} we also list the median value and $1\sigma$ uncertainty for each (sub)sample, with all data adopted from the NSA catalogue. We can see that the mid-IR sample has the lowest mass, with a median value of $10^{8.7}~M_\odot$, and the bluest $g-r$ color, with a median value of 0.26. Among these samples, a significantly redder $g-r$ color is found for the optical sample (R13), providing clear evidence of the selection effect introduced by the BPT method.

\subsection{Stellar Population Synthesis}

In this study, we use STARLIGHT \citep[hereafter C05]{Cid2005}, a stellar population synthesis code that combines empirical population synthesis with evolutionary synthesis models for our spectra fitting. STARLIGHT fits an observed spectrum with a linear combination of $N_{\star}$ simple stellar populations (SSPs) and returns the fractional contribution of $j$th SSP $N_{\star j}$ to the total synthetic flux at the normalization wavelength $\lambda _0$, parameterized by the population vector $\bm{x}$. Our spectra take from SDSS DR8 \citep{Aihara2011}, which has a resolution of $\lambda / \Delta \lambda \sim 1800$ and spectra range 3800-9200 $\text{\AA}$. Before the fitting, The Milky Way foreground extinction is corrected using the extinction law of \citet{Cardelli1989} and \citet{O'Donnell1994}, assuming $R_V=3.1$ and $A_V$ is derived from \citet{Schlegel1998} in NASA/IPAC Extragalactic Database (NED), spectra also shift to the rest frame and resample in steps of 1 $\text{\AA}$.

Following the similar configuration of \citet{Zhao2011} and C20, we mask obvious emission lines and give double weight to Ca\,{\sc ii}\,K\,$\lambda3934$ and Ca\,{\sc ii} triplets which are among the strongest stellar absorption features less affected by nearby emission lines. The observed spectra are normalized at the median value of flux between 4010 $\text{\AA}$ and 4060 $\text{\AA}$ and adopted 4730-4780 $\text{\AA}$ as the signal-to-noise ratio ($\mathrm{S/N}$) window. All the SSPs normalized at $\lambda _0 = 4020 \text{\AA}$. We use a template base with $N_{\star}=100$ SSPs covering 25 ages from 1 Myr to 18 Gyr and four metallicities Z=0.0001, 0.0004, 0.004 and 0.008 from \citet{Bruzual2003}. In addition, a power-law (PL) component with an index of $\alpha=-1.5$ is added to represent an AGN featureless continuum (FC) $F_\nu \propto \nu^{\alpha}$, the flux contribution fraction of this nonstellar component at $\lambda _0 = 4020 \text{\AA}$ is denoted by $x_{\rm {AGN}}$. Meanwhile, the samples presented in this paper show strong star formation, implying that the galaxy light may be dominated by star-forming regions of host galaxies, therefore, a pure stellar population model without a PL component is also carried out. The difference between the pure stellar population model and a PL component-joined AGN model will be discussed in Section \ref{sec:Stellar Population Properties from the Pure Stellar Population Model}. 

The galactic intrinsic extinction $A_{V,\star}$ is modeled as due to foreground dust screen with the law of \citet{Calzetti1994} and $R_V=4.05$ \citep{Calzetti2000}. It should be noted that the $A_{V,\star}$ does not require to be positive during the fitting, the reasons discussed in C05 and \citet{Mateus2006}. If $A_{V,\star}<0$, we rerun the program with a constraint of $A_{V,\star} \geqslant 0$, then compare the results. If light or mass-weight mean stellar age changed over 0.3 dex or stellar population contribution fraction changed over 20\%, we adopt the re-fitted result. For our samples, 14 out of 25 and 13 out of 18 sources are replaced in the following analysis for synthesis of the AGN-joined model and pure stellar population model, respectively. 

\begin{figure}[tb]
\centering
    \includegraphics[width=0.47\textwidth, angle=0]{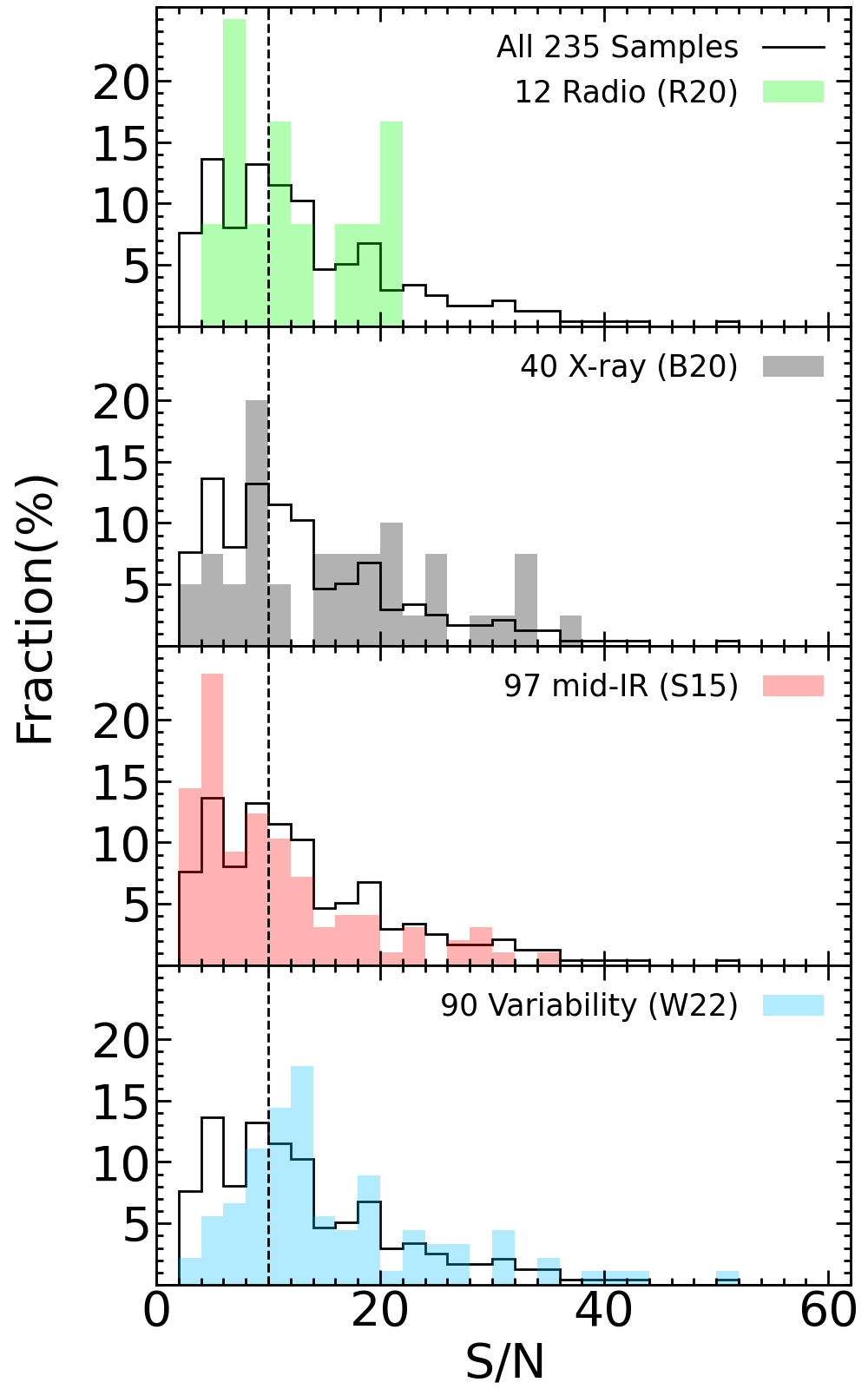}
    \caption{Distribution of $\mathrm{S/N}$ for various samples. The dash line indicates $\mathrm{S/N}=10$.}
    \label{fig_s1}
\end{figure}

As shown in C05, \citet{Mateus2006}, \citet{Ge2018} and \citet{Woo2024}, the uncertainties of the synthesis results depend on the $\mathrm{S/N}$ ratio of the input spectrum. For the whole sample, the $\mathrm{S/N}$ varies between 2.5 and 51.5, with a median value of $11.1\pm7.9$, and the distribution of $\mathrm{S/N}$ is shown in Fig. \ref{fig_s1}. For each subsample, the median $\mathrm{S/N}$ ratios are $10.2\pm5.2$ (radio), $16.2\pm10.5$ (X-ray), $8.1\pm5.6$ (mid-IR) and $12.8\pm6.6$ (variability), indicating that in general the derived stellar age, metallicity (see \S \ref{sec:Stellar Population Properties}) and stellar extinction ($E(B-V)$) would have uncertainties of $\lesssim0.15$ dex, $\lesssim0.15$ dex and 0.05 dex, respectively. We also note that there are about 14.9\% (2.6\%) of our sources having $\mathrm{S/N}<5$ ($<3$), which may suffer a relatively lage uncertainty, i.e., up to 0.3 dex, 0.4 dex and 0.1 dex in the stellar age, metallicity and stellar extinction, respectively (e.g., \citealt{Mateus2006,Woo2024}).

\section{Result and Discussion} \label{sec:Result and Discussion}
\subsection{Pure-Emission Spectra and the BPT Diagram}\label{sec:bpt}
\begin{figure}[tb]
\centering
    \includegraphics[width=0.47\textwidth, angle=0]{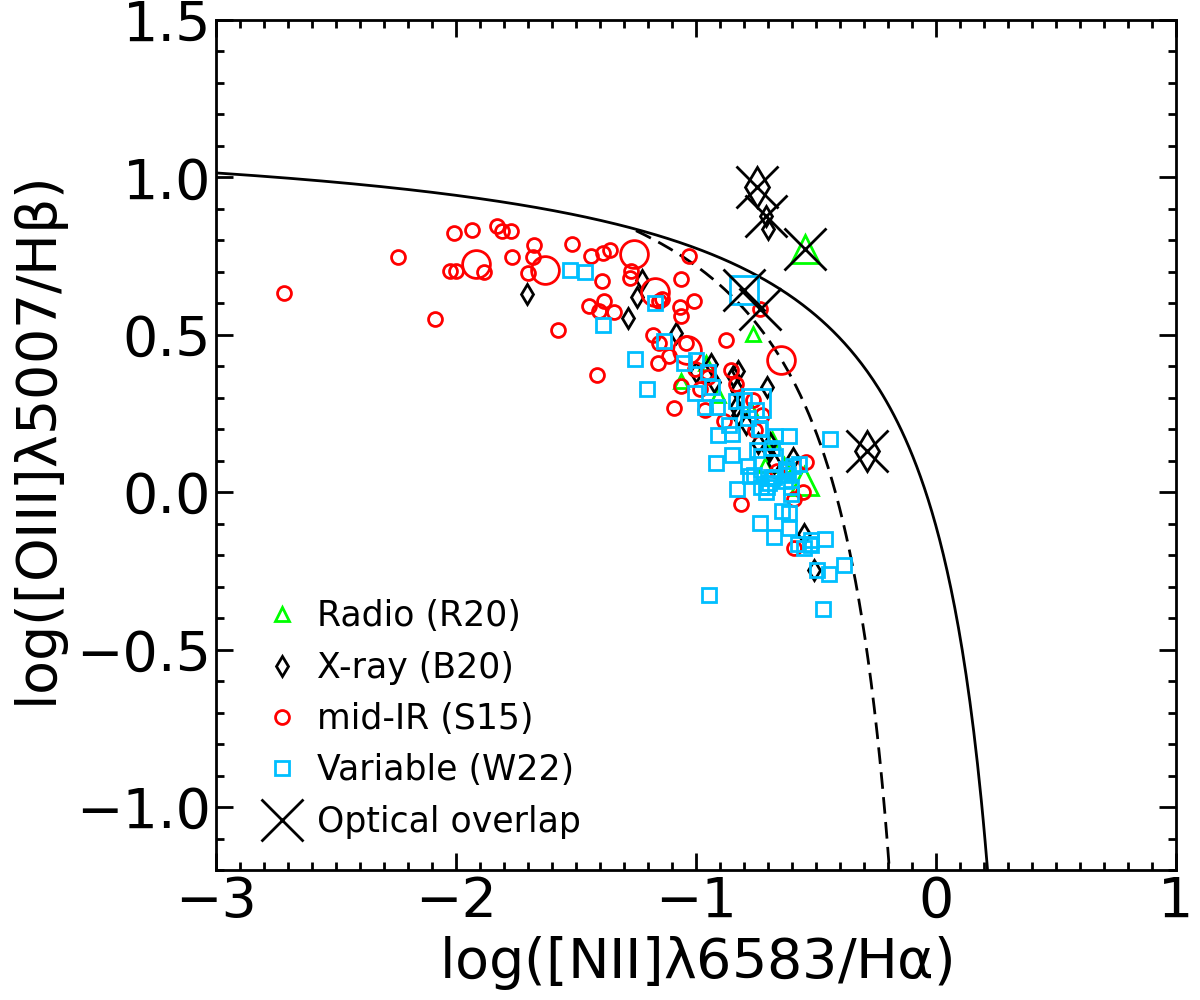}
    \caption{\OIIII /\Hb\ versus \NII / \Ha\ BPT diagram for the 171 sources having ${\mathrm{S/N}} >3$ for all of the four emission lines. Radio-, X-ray-, mid-IR- and variability-selected objects are represented by (lime) triangles, (black) diamonds, (red) circles and (light blue) squares, respectively. The six sources with an additional crosses are also included in the R13 optical sample. Larger symbols demonstrate the sources showing a broad component in \Ha. The sample is vastly overwhelmed (162/171) by pure star-forming galaxies defined by \citet[dashed line]{Kauffmann2003}, and only 4 sources lie above the theoretical maximum starburst line (solid line) defined by \citet{Kewley2001a}.}
    \label{fig2}
\end{figure}

We measure line fluxes, such as \Hb, \OIIII, \Ha\, and \NII, using the pure-emission spectrum, which is obtained by subtracting the synthesized continuum and stellar absorption spectrum returned by STARLIGHT. The fitting is done using a single Gaussian model, then we compare the reduced $\chi ^2$ value with a two-component Gaussian model for \Hb, \Ha, and \NII. If the $\chi ^2$ value of the two-component model is closer to 1 and improved by 20\% than the single Gaussian model, the result of the two-component Gaussian model is accepted. We refer to the object with a broad component of Full Width at Half Maximum (FWHM) at least 500 km/s in two Gaussian models as a source with broad line and calculate the line flux after removing the broad component. For most of majority galaxies, a single Gaussian is fitted well, only 14 sources require a two Gaussian model for \Ha. 

Finally, there are 171 sources (10 radio, 26 X-ray, 71 mid-IR (two in the radio, and two in the variability), 68 variability) having $\mathrm{S/N}>3$ for all of the four emission lines mentioned above. In Fig. \ref{fig2}  we plot the \OIIII/\Hb\ versus \NII/\Ha\ diagram (\citealt{Baldwin1981,Veilleux1987}). Obviously, the vast majority (95\%) of the sample sources are located in the region classified as star formation, and only 4 sources fall in the AGN-dominated region and 5 in the composite region, and such a small fraction of AGNs classified by the BPT method has already been noted in the original works that selected these AGN-host dwarfs. 

One possible reason is that the emission lines might be dominated by star-formation in the host galaxies, for which most the radiation from the central AGN should be diluted heavily. It is also likely that the low-mass BHs in dwarf galaxies might produce a harder radiation field, as predicted by the theoretical models in \citet{2019ApJ...870L...2C}, resulting in a net decrease in the  \OIIII/\Hb\ and \NII/\Ha\ emission line ratios. Furthermore, the low-metallicity of dwarf galaxies will also lead the observed line ratios to moving further into the star-forming region of the diagram.
\begin{table}
\bc
\begin{minipage}[]{140mm}
\caption[]{Median Values and Dispersion of the Derived Stellar Population Properties\label{tab2}}\end{minipage}
\setlength{\tabcolsep}{6pt}
\small
 \begin{tabular}{ccccccccccccc}
  \hline\noalign{\smallskip}
Sample & $\langle \log t_\star \rangle_L$ & $\langle \log t_ \star \rangle_M$ & $ \sigma_{L}(\log t_{\star})$ & $\sigma_{M}(\log t_{\star})$ & $\langle Z_\star\rangle_L/Z_\odot$ & $\langle Z_\star\rangle_M/Z_\odot$\\
  \hline\noalign{\smallskip}
All & 8.19$\pm$0.95 & 9.67$\pm$0.34 & 1.06$\pm$0.32 & 0.51$\pm$0.21 & 0.20$\pm$0.08 & 0.24$\pm$0.14 \\
radio (R20) & 8.01$\pm$0.76 & 9.79$\pm$0.14 & 1.14$\pm$0.36 & 0.57$\pm$0.18 & 0.20$\pm$0.07 & 0.24$\pm$0.07 \\
X-ray (B20) & 8.28$\pm$0.84 &  9.66$\pm$0.33 & 1.08$\pm$0.23 & 0.54$\pm$0.17 & 0.23$\pm$0.09 & 0.25$\pm$0.13 \\
mid-IR (S15) & 7.74$\pm$1.14 & 9.67$\pm$0.41 & 1.01$\pm$0.44 & 0.53$\pm$0.30 & 0.17$\pm$0.07 & 0.19$\pm$0.16 \\
variability (W22) & 8.44$\pm$0.81 & 9.68$\pm$0.27 & 1.07$\pm$0.30 & 0.47$\pm$0.16 & 0.23$\pm$0.07 & 0.26$\pm$0.12 \\
optical (C20) & 8.88$\pm$0.69 & 9.80$\pm$0.14 & 1.12$\pm$0.31 & 0.36$\pm$0.07 & 0.29$\pm$0.05 & 0.30$\pm$0.08 \\
$f_{r}<20\%$ (S1) & 8.31$\pm$0.91 & 9.66$\pm$0.36 & 1.06$\pm$0.35 & 0.47$\pm$0.19 &  0.21$\pm$0.08 & 0.24$\pm$0.16 \\
$f_{r} \geq 20\%$ (S2) & 7.91$\pm$1.10 & 9.69$\pm$0.32 & 1.08$\pm$0.29 & 0.58$\pm$0.23 & 0.19$\pm$0.08 & 0.22$\pm$0.11 \\
$f_{r}<f_{r,\,{\rm med}} \, ^a$ & 8.42$\pm$1.04 & 9.67$\pm$0.35 & 1.06$\pm$0.31 & 0.47$\pm$0.20 & 0.21$\pm$0.09 & 0.24$\pm$0.16 \\
$f_{r} \geq f_{r,\,{\rm med}}$ & 8.01$\pm$0.95 & 9.68$\pm$0.32 & 1.08$\pm$0.33 & 0.57$\pm$0.20 & 0.20$\pm$0.07 & 0.23$\pm$0.12 \\
$x_{\rm {AGN}}=0$ (S3) & 8.14$\pm$0.86 & 9.82$\pm$0.33 & 1.20$\pm$0.31 & 0.48$\pm$0.24 & 0.17$\pm$0.07 & 0.16$\pm$0.12 \\
$x_{\rm {AGN}}>0$ (S4) & 8.20$\pm$0.95 & 9.66$\pm$0.33 & 1.03$\pm$0.33 & 0.51$\pm$0.21 & 0.22$\pm$0.08 & 0.25$\pm$0.13 \\
  \noalign{\smallskip}\hline
\end{tabular}
\ec
\tablecomments{0.86\textwidth}{$^af_{r,\,{\rm med}}(=13.6\%)$ is median value of $f_r$ for all samples}
\end{table}
\subsection{Stellar Population Properties} \label{sec:Stellar Population Properties}

\begin{figure}
\centering
    \includegraphics[width=0.95\textwidth, angle=0]{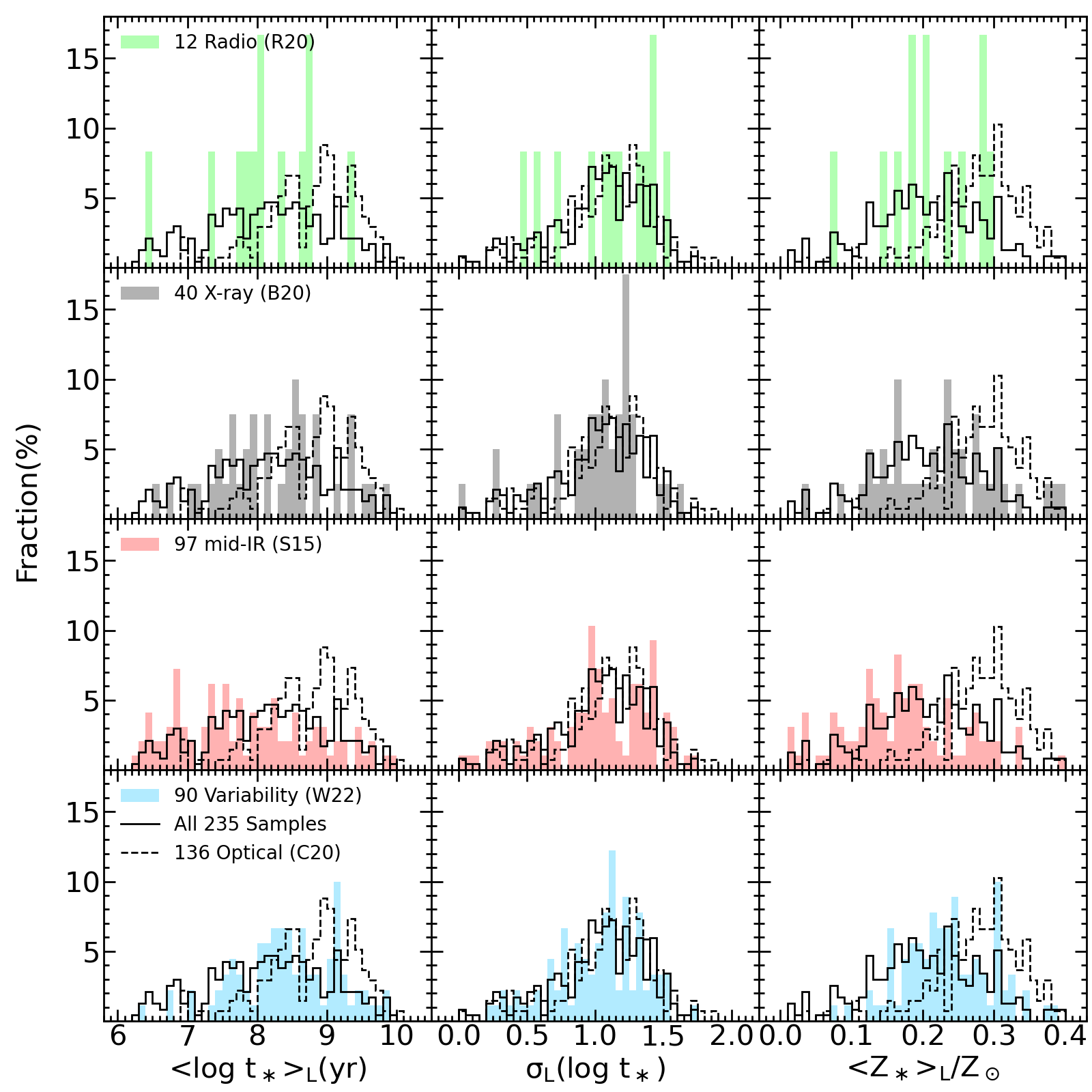}
    \caption{Fraction distributions of $\langle \log t_{\star}\rangle_{L}$ (left), $\sigma_{L}(\log t_{\star})$ (middle) and $\langle Z_{\star}\rangle_{L}$ (right) for different samples represented by different colored histograms as annotated. In each panel, the whole (black solid line) and the optical (black dashed line) samples are also plotted for comparison.
\label{fig3}}
\end{figure}

\subsubsection{Mean Stellar Age}
Following the methods of C05, we calculate the light- and mass-weighted mean stellar ages respectively using $\langle \log t_{\star}\rangle_{L}=\sum\limits_{j=1}^{N_{\star}} x_{j}\log t_{j}$ and
$\langle \log t_{\star}\rangle_{M}=\sum\limits_{j=1}^{N_{\star}} \mu_{j}\log t_{j}$,
where $x_{j}$ and $\mu_{j}$ are fractional contributions of the $j$th SSP with age $t_{j}$ in stellar luminosity and mass, respectively, $N_{\star}$ represents the total number of SSPs. The light-weighted mean stellar age reveals the SFH due to the direct relation with the observed spectrum and reflects the epoch of bright and massive star formation, the mass-weighted mean stellar age is more physical and associated with the age of the stellar population dominating galactic mass. 
Similarly, it is convenient to define the light (mass)-weighted standard deviation of the log age, $\sigma_{L}(\log t_{\star})$ ($\sigma_{M}(\log t_{\star})$) to differentiate a single population from diverse SFHs, namely,
$\sigma_{L}(\log t_{\star})=\left[ \sum\limits_{j=1}^{N_{\star}} x_{j}(\log t_{j}-\langle \log t_{\star}\rangle_{L})^2\right ]^{1/2}$ 
and
$\sigma_{M}(\log t_{\star})=\left [ \sum\limits_{j=1}^{N_{\star}} \mu_{j}(\log t_{j}-\langle \log t_{\star}\rangle_{M})^2\right ] ^{1/2}$.

Fig. \ref{fig3} plots the distributions of $\langle \log t_{\star}\rangle_{L}$ (left) and  $\sigma_{L}(\log t_{\star})$ (middle) for different samples. The light-weighted mean stellar age $\langle \log t_{\star}\rangle_{L}$ of all samples in our work is in the range of $10^6$-$10^{10}$ yr. Among the four samples, the stellar population properties of the variability sample are most similar to that of the X-ray sample. The mid-IR sample is the youngest, with a median values of $\langle \log t_{\star}\rangle_{L}= 7.74$ (see Table \ref{tab2}), and shows a wider span of $\langle \log t_{\star}\rangle_{L}$ than the other three samples, in the sense that the youngest ($\langle \log t_{\star}\rangle_{L} = 6.24$) and oldest ($\langle \log t_{\star}\rangle_{L} = 9.94$) galaxies come both from the mid-IR sample. The fractions of galaxies having $\langle \log t_{\star}\rangle_{L}<10^8$ yr ($\langle \log t_{\star}\rangle_{L}>10^9$ yr) are 41.7\% (8.3\%), 42.5\% (17.5\%), 57.7\% (13.4\%) and  23.3\% (27.8\%), respectively, for the radio, X-ray, mid-IR and variability samples. As we can see, the variability sample has the smallest (but largest) fraction of galaxies of $\langle \log t_{\star}\rangle_{L}<10^8$ yr ($\langle \log t_{\star}\rangle_{L}>10^9$ yr), and thus it is the oldest sample with the median value of $\langle \log t_{\star}\rangle_{L} = 8.44$, about 0.7 dex older than the mid-IR sample. Whereas for the radio and X-ray samples, the median values of $\langle \log t_{\star}\rangle_{L}$ are 8.01 and 8.28, respectively.

For comparison, we also plot the distributions of $\langle \log t_{\star} \rangle_{L}$ and $\sigma_{L}(\log t_{\star})$ for the optical sample in Fig. \ref{fig3} using dashed-line histograms. As shown in the figure, the optical sample is generally older the other four samples, with $\langle \log t_{\star}\rangle_{L}$ peaking at $\sim$9. Furthermore, it also has a narrower span of $\langle \log t_{\star}\rangle_{L}$, and only $\sim$11\% of the sample galaxies have $\langle \log t_{\star}\rangle_{L}<8$. However, all but the radio sample, which has a very limited size, show similar distributions in $\sigma_{L}(\log t_{\star})$, suggesting complex and diverse SFHs in all of the AGN-host dwarf galaxies, irrespective of the selection method.

A recent study of the stellar population of 26 Changing-look (CL) AGNs by \citet{Jinwu2022} shows that the CL-AGNs lie between starburst and normal galaxies, and most (88.5\%) have $\langle \log t_{\star}\rangle_{L}>9$, older than any dwarf sample. In \citet{Mahoro2022}, however, 77\% of the 17 X-ray selected, far-IR detected AGN-host galaxies have $\langle \log t_{\star}\rangle_{L}<9$, and the median age of this sample is $\langle \log t_{\star}\rangle_{L}=8.5$, only $\sim$0.2 dex older than our X-ray sample. Therefore, the optically-selected AGN-hosts, irrespective of stellar mass, are generally have older stellar populations.

\begin{figure}
\centering
    \includegraphics[width=0.47\textwidth, angle=0]{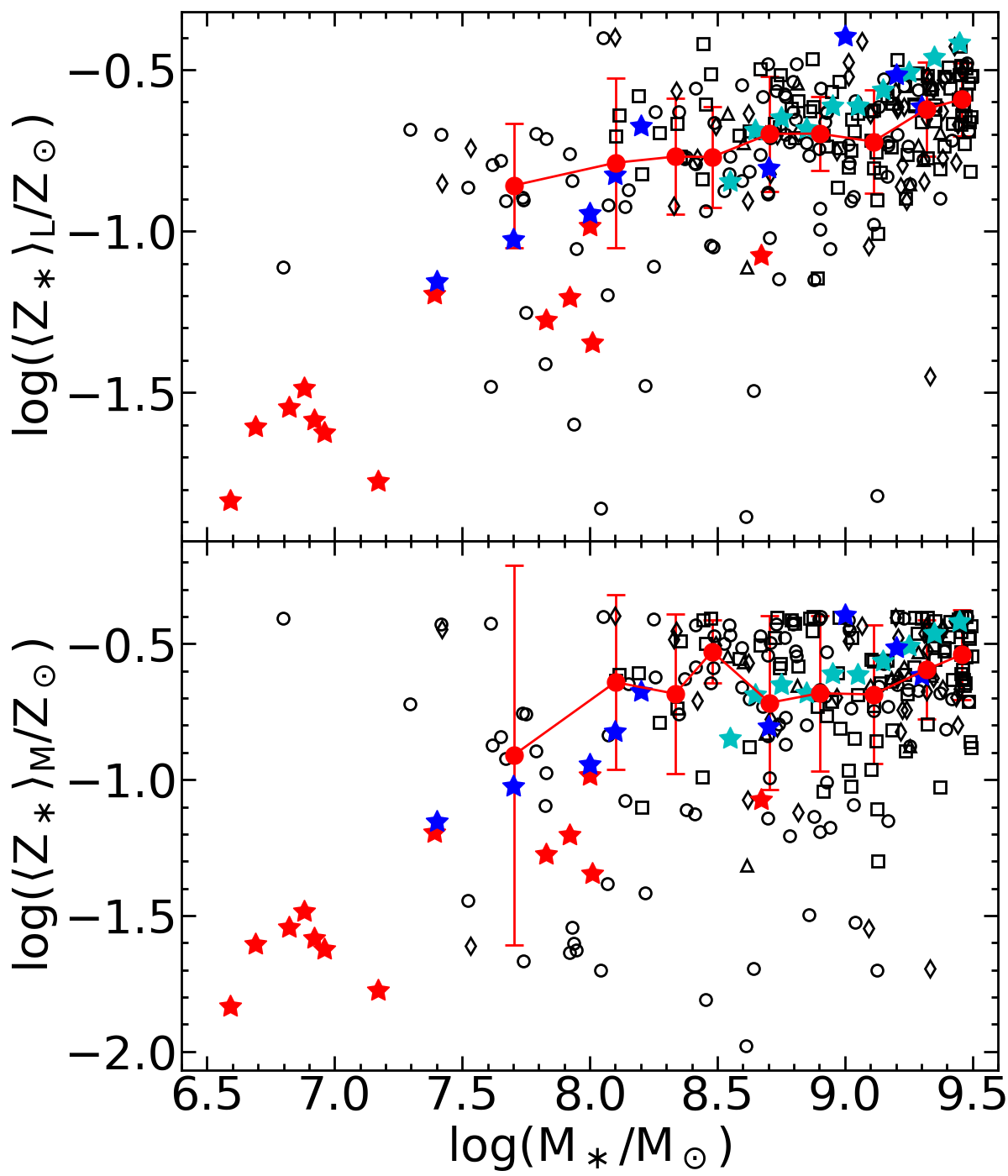}
    \caption{Mass-metallicity relation (MZR) for light-weighted mean stellar metalicity $\langle Z_{\star}\rangle_{L}$ (top panel) and mass-weighted mean stellar metalicity $\langle Z_{\star}\rangle_{M}$ (bottom panel). Black open symbols represent all AGN-host dwarf galaxy samples, as in Fig. \ref{fig2}. Red solid circles show the median values of $\langle Z_{\star}\rangle$ in each mass bin with error bars giving the dispersion. For comparison, we also plot the metallicities for star-forming galaxies \citep[cyan stars]{Zahid2017}, those nearby galaxies measured with spectroscopy of individual blue supergiant stars \citep[and references therein, blue stars]{Davies2017} and dwarf galaxies in the Local Group \citep[red stars]{Kirby2013}. All metallicities have been scaled to the solar metallicity ($Z_\odot$ = 0.02) adopted in this work.
\label{fig4}}
\end{figure}

\subsubsection{Mean Stellar Metallicity}
For AGN-host galaxies, it is very difficult (or even impossible) to determine the nebular metallicity due to the AGN contamination of emission lines. Therefore, the stellar metallicity obtained through stellar population synthesis is specially useful for the study on the chemical properties of the AGN-host dwarfs (e.g., C20). To this end, we calculate the light- and mass-weighted mean stellar metallicities with a similar way to the mean stellar ages, namely,
$\langle Z_{\star}\rangle_{L}=\sum\limits_{j=1}^{N_{\star}} x_{j}Z_{j}$
and
$\langle Z_{\star}\rangle_{M}=\sum\limits_{j=1}^{N_{\star}} \mu_{j}Z_{j}$. The distributions of $\langle Z_{\star}\rangle_{L}$ for each sample are plotted in the right panel of Fig. \ref{fig3}. We also list the median values of $\langle Z_{\star}\rangle_{L}$ and $\langle Z_{\star}\rangle_{M}$ in Table \ref{tab2}.

Among the four samples, the mid-IR sample is most metal-poor, with a median value of $\langle Z_{\star}\rangle_{L}= 0.17~Z_\odot$, and the variability sample is most metal-rich, with a median value of $\langle Z_{\star}\rangle_{L}= 0.23~Z_\odot$. When considering the optical sample, it has the largest median $\langle Z_{\star}\rangle_{L}$ of 0.29~$Z_\odot$, and about 90\% of the optically-selected galaxies have $\langle Z_{\star}\rangle_{L}>0.2~Z_\odot$, as shown in the bottom right panel of Fig. \ref{fig3} (dashed-line histogram). These differences can be explained with the mass-metallicity relation (MZR; e.g., \citealt{Zahid2017}; C20), as the mid-IR and optical samples have the lowest and highest (medidan) stellar masses, respectively, as listed in Table \ref{tab1}. 

To further check the MZR for our sample galaxies, we plot stellar mass versus the light- (upper) and mass-weighted (bottom) metallicites in Fig. \ref{fig4}, with (red) solid circles representing the median values of each mass bin. We also overplot the star-forming galaxies in \citet[][cyan stars]{Zahid2017}, for which the stellar metallicities are also obtained using stellar population synthesis, and nearby galaxies in \citet[][red stars]{Kirby2013} and \citet[][blue stars]{Davies2017}. Compared with the optical sample in C20, our dwarfs extend to a much smaller ($\sim$4.5$\times$) mass. As shown in the figure, for our AGN-host dwarfs there exists a similar MZR to normal dwarf galaxies. Our result confirms the finding in C20 and provides further evidence for that AGNs are unlikely to strongly influence the chemical evolution of their host dwarf galaxies.

On the other hand, we can also see from Fig. \ref{fig4} that there is a large spread in the plot, especially for the mid-IR sample (open circles). This raises a tricky problem of the age-metallicity degeneracy, which confuses old metal-poor systems with young metal-rich ones, and thus would lead to an obvious deviation from the MZR. After investigating the morphologies of the sources significantly deviating from the MZR, we find that those having low masses but high metallicities tend to be early-type (S\'{e}rsic index $n>3$) galaxies with young $\langle \log t_{\star}\rangle_{L}$, while those with large masses but low metallicities tend to be late-type (S\'{e}rsic index $n<3$) galaxies with old $\langle \log t_{\star}\rangle_{L}$. This result is opposite to the general finding that late-type galaxies have younger $\langle \log t_{\star}\rangle_{L}$, providing another evidence for that the age-metallicity degeneracy should have severely affected a small part of our sample galaxies.

\begin{figure}
\centering
    \includegraphics[width=0.95\textwidth, angle=0]{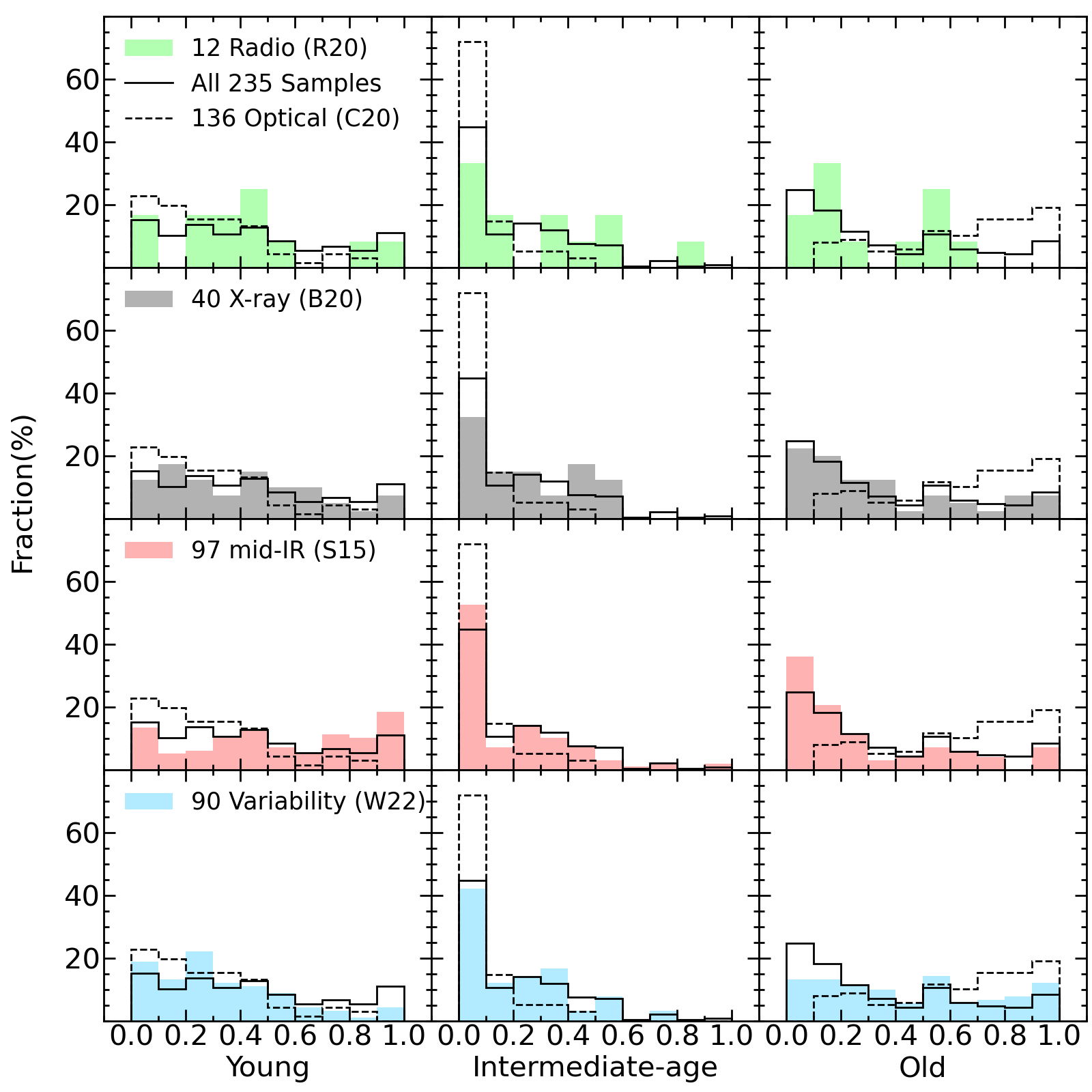}
    \caption{Fractional distributions of luminosity fractions of young ($t< 10^8$ yr), intermediate-age ($10^8 < t < 10^9$ yr), and old ($t > 10^9$ yr) stellar populations for different samples. In each panel, the black solid- and dashed-line histograms represent all of the 235 AGN-host dwarf galaxies and the optical sample, respectively.
\label{fig5}}
\end{figure}

\subsubsection{Star Formation Histories}
In order to reduce the uncertainties in the individual components (C05), and to obtain a general view of the SFHs of these AGN-host dwarfs, we arrange their stellar populations into three groups, namely, young ($t< 10^8$yr), intermediate-age ($10^8 < t < 10^9$ yr), and old ($t > 10^9$ yr) populations, denoted by $x_{\rm Y}$ ($\mu_{\rm Y}$), $x_{\rm I}$ ($\mu_{\rm I}$) and $x_{\rm O}$ ($\mu_{\rm O}$) respectively, in line with previous studies (e.g., C05; C20). For $\mathrm{S/N}\geqslant10$, the uncertainties can be $\Delta x_{\rm Y} \leq 0.05$, $\Delta x_{\rm I} \leq 0.1$ and $\Delta x_{\rm O}\leq 0.1$ (C05).
Fig. \ref{fig5} shows the distributions of luminosity contribution fractions in different population groups. Compared with the optical sample, for which the fraction increases with decreasing $x_{\rm Y}$, our whole sample has a flatter distribution in $x_{\rm Y}$. Furthermore, the trend of $x_{\rm O}$ for the optical sample is opposite to our whole sample, which generally shows a smaller fraction as $x_{\rm O}$ increasing. 

For each our sample, 25.0\% (radio), 35.0\% (X-Ray), 52.6\% (mid-IR) and 22.2\% (variability) of the galaxies have their $x_{\rm Y}$ larger than $50\%$.  
Generally, a much smaller fraction of the galaxies have $x_{\rm I}>50\%$, i.e., 25.0\%, 12.5\%, 8.2\% and 11.1\% respectively for the corresponding samples. In the radio, X-ray and variability samples, we also find that a small fraction ($<10\%$) of the sources have extremely large $x_{\rm Y}$ ($>90\%$). While for the mid-IR sample, a much larger fraction ($\sim$20\%) of the member galaxies have $x_{\rm Y}>90\%$, and the young population dominates the luminosity. 

Although the variability sample has the smallest contribution from the young population,  the fraction is still larger than that of the optical sample, for which only 18 (13.2\%) sources have $x_{\rm Y}>50\%$ and no source shows $x_{\rm I}>50\%$. The most significant difference is between the contributions from the old population: 97 (71.3\%) out of 136 sources of the optical sample have $x_{\rm O} > 50\%$, whereas only 33.3\%, 30.0\%, 24.7\% and 46.7\% objects, respectively for the radio, X-ray, mid-IR and variability samples, have $x_{\rm O} > 50\%$. These results indicate that most of our sample galaxies should have experienced star-forming activities in the recent 1 Gyr.

Furthermore, C20 shows that for the optically-selected AGN-host dwarfs with stellar mass of $10^9-3\times10^{9}\,M_\odot$, only 4\% have $\mu_{\rm Y}>5\%$, far less than the fraction of 25-40\% found by \citet{Kauffmann2014} for the same mass range. The authors suggest that the much lower fraction may be caused by the suppression of AGNs in the current star formation of the host galaxies, and/or the selection bias inherent to the optical sample. For our galaxies within the same mass range, the fractions having $\mu_{\rm Y}>5\%$ are 14.3\%, 14.8\%, 9.1\%, and 5.8\%, respectively, for the radio, X-ray, mid-IR and variability samples, resulting in a final fraction of 9.5\% (10/105) for the whole sample. Our results, which unlikely suffer from the selection effect, further suggest that AGNs in dwarfs likely suppress the current star formation of the host galaxy, consistent with the finding in \citet{Cai2021} that AGN-host dwarfs have an older light-weighted age than the control sample. 

\begin{figure}[tb]
\centering
    \includegraphics[width=0.47\textwidth, angle=0]{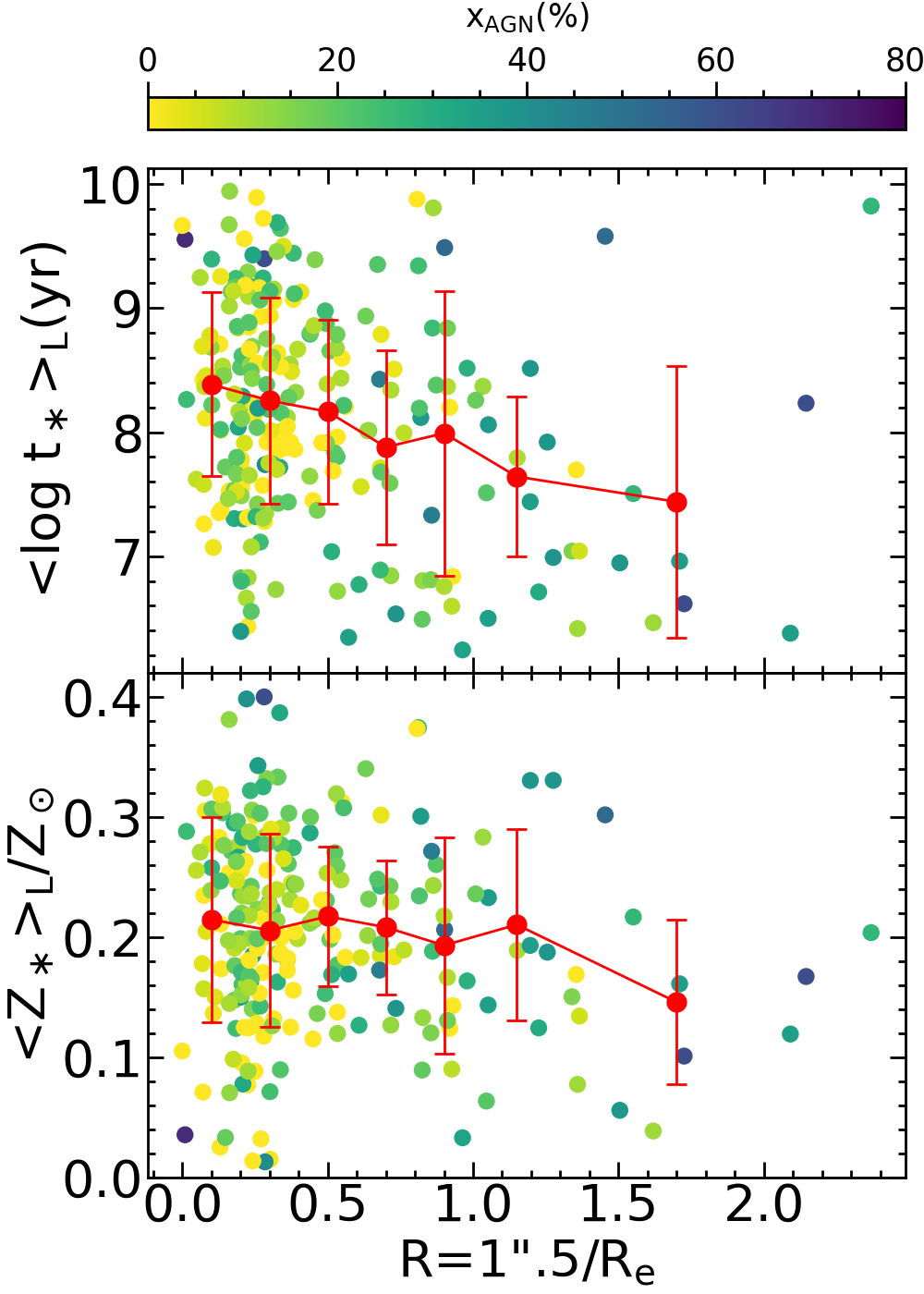}
    \caption{Distributions of $\langle \log t_{\star}\rangle_{L}$ (top) and $\langle Z_{\star}\rangle_{L}$ (bottom) as a function of the normalized radius $R=1\arcsec.5/R_e$ for all of the 235 AGN-host dwarf galaxies, colored with $x_{\rm {AGN}}$. The error bars in the y-axis show the dispersion of the mean value in each $R$ bin.
\label{fig6}}
\end{figure}

\subsubsection{Pseudo-Radial Profile}

The spectra used in this work are obtained from fixed-size $3\arcsec$-diameter fibers of SDSS, which cover different physical scales and light fractions of the entire galaxy, depending on the distance of the target. For our whole sample, it corresponds to a physical size of about 64 pc and 3.3 kpc for the closest and farthest sources, respectively. Whereas the light fraction, $f_r (\equiv 100 \times 10^{-0.4(m_{\rm fiber}-m_{\rm Petro})})$, which describes the covering fraction of the fixed-size aperture and is calculated with the fiber and total galaxy magnitudes in the $r$ band following \citet{Zhao2011}, is in the range of $\sim$$0.6-100\%$, and has a median value of $f_{r,\,{\rm med}}=13.6\%$. Therefore, more than a half of our sample galaxies have $f_r$ less than 20\%, which is required to minimize the aperture effect as demonstrated in \citet{Kewley2005}. Indeed, there is a significant difference (0.4 dex) between the median $\langle \log t_{\star}\rangle_{L}$ for the two subsamples (S1 and S2) divided with $f_{r}=20\%$, as shown in Table \ref{tab2} (also see Table \ref{tab3}), and the subsample (S2) with a larger $f_r$ has a younger light-weighted stellar age, suggesting that the inner part might be older.

In order to overcome the limitation caused by the fixed-size aperture, and further investigate the AGN influence on the host galaxy, we define a normalized radius, namely $R=1\arcsec.5/R_e$, where $R_e$ is the radius enclosing 50 percent of the galaxy light in  $r$ band. In this way, we can alleviate the aperture effect and obtain a pseudo-radial profile of the stellar populations, as shown in Fig. \ref{fig6}. It can be seen from the top panel that there exists a negative gradient in the MEAN light-weighted stellar age for $R\lesssim1$, similar to previous works for normal galaxies \citep{Gonzalez2015,Li2015,Morelli2015,Roig2015,Zheng2017}, which is implying an inside-out formation scenario. However, \citet{Cai2021} find a flat radial-profile of $\langle \log t_{\star}\rangle_{L}$ for their optically-selected AGN-host dwarfs, suggesting that the AGN effect seems to result in an overall quenching of the host galaxy. We speculate that the AGN-host dwarfs in the current work and \citet{Cai2021} are at different evolution stages, in the sense that our sample galaxies have abundant cold gas to fuel the AGN and star formation (early stage), while the sources in \citet{Cai2021} have little gas to support the star formation (late stage). It needs more IFU data of AGN-hosts selected by methods other than the BPT diagram to obtain a solid conclusion.

Many works have reported a negative metallicity gradient, and this trend becomes flatter and even disappears for low-mass systems \citep{Zheng2017,Belfiore2018,Cai2021}. As illustrated in the bottom panel of Fig. \ref{fig6}, we do not see any obvious trend in $\langle Z_{\star}\rangle_{L}$ within 1$R_e$ for our whole sample, and the apparent decrease at larger radii is believed to be caused by the much fewer data points we used.
Our result is generally consistent with the finding in \citet{Cai2021}, who show that AGN-hosts and normal dwarfs have very flat radial profiles of $\langle Z_{\star}\rangle_{L}$, with a tiny gradients of $-0.03$ and $-0.06$ dex/$R_e$, respectively, for the inner region of $1R_e$. These results further confirm that AGNs have no strong impact on the chemical evolution of their host galaxies.

\subsection{Stellar Population Properties from the Pure Stellar Population Model} \label{sec:Stellar Population Properties from the Pure Stellar Population Model}
\begin{table}[tb]
\bc
\begin{minipage}[]{140mm}
\caption[]{Median Values and Dispersion of the Derived Stellar Population Properties With Pure Stellar Population Model\label{tab3}}\end{minipage}
\setlength{\tabcolsep}{6pt}
\small
 \begin{tabular}{ccccccccccccc}
  \hline\noalign{\smallskip}
Sample & $\langle \log t_\star \rangle_L$ & $\langle \log t_ \star \rangle_M$ & $ \sigma_{L}(\log t_{\star})$ & $\sigma_{M}(\log t_{\star})$ & $\langle Z_\star\rangle_L/Z_\odot$ & $\langle Z_\star\rangle_M/Z_\odot$\\
  \hline\noalign{\smallskip}
All & 7.99$\pm$0.80 & 9.74$\pm$0.26 & 1.27$\pm$0.23 & 0.55$\pm$0.22 & 0.18$\pm$0.06 & 0.19$\pm$0.11 \\
radio (R20) & 7.83$\pm$0.65 & 9.73$\pm$0.14 & 1.38$\pm$0.16 & 0.61$\pm$0.23 & 0.17$\pm$0.04 & 0.20$\pm$0.06 \\
X-ray (B20) & 8.04$\pm$0.66 & 9.78$\pm$0.20 & 1.23$\pm$0.18 & 0.60$\pm$0.26 & 0.20$\pm$0.07 & 0.20$\pm$0.09 \\
mid-IR (S15) & 7.60$\pm$0.98 & 9.72$\pm$0.35 & 1.29$\pm$0.26 & 0.62$\pm$0.27 & 0.14$\pm$0.06 & 0.15$\pm$0.10 \\
variability (W22) & 8.32$\pm$0.61 & 9.75$\pm$0.21 & 1.26$\pm$0.20 & 0.50$\pm$0.17 & 0.20$\pm$0.05 & 0.24$\pm$0.11 \\
$f_{r}<20\%$ (S1) & 8.12$\pm$0.78 & 9.72$\pm$0.26 & 1.26$\pm$0.22 & 0.51$\pm$0.19 & 0.19$\pm$0.07 & 0.21$\pm$0.14 \\
$f_{r} \geq 20\%$ (S2) & 7.69$\pm$1.05 & 9.78$\pm$0.21 & 1.32$\pm$0.20 & 0.64$\pm$0.24 & 0.17$\pm$0.05 & 0.18$\pm$0.08 \\
$f_{r}<f_{r,\,{\rm med}}$ & 8.19$\pm$0.88 & 9.74$\pm$0.25 & 1.26$\pm$0.22 & 0.49$\pm$0.19 & 0.18$\pm$0.07 & 0.21$\pm$0.14 \\
$f_{r} \geq f_{r,\,{\rm med}}$ & 7.88$\pm$0.90 & 9.76$\pm$0.25 & 1.29$\pm$0.21 & 0.62$\pm$0.23 & 0.17$\pm$0.06 & 0.18$\pm$0.09 \\
S3 & 8.10$\pm$0.87 & 9.78$\pm$0.33 & 1.15$\pm$0.27 & 0.51$\pm$0.21 & 0.17$\pm$0.08 & 0.17$\pm$0.14 \\
S4 & 7.96$\pm$0.81 & 9.74$\pm$0.24 & 1.29$\pm$0.21 & 0.56$\pm$0.21 & 0.18$\pm$0.06 & 0.20$\pm$0.10 \\
  \noalign{\smallskip}\hline
\end{tabular}
\ec
\end{table}

\begin{figure}
\centering
    \includegraphics[width=0.47\textwidth, angle=0]{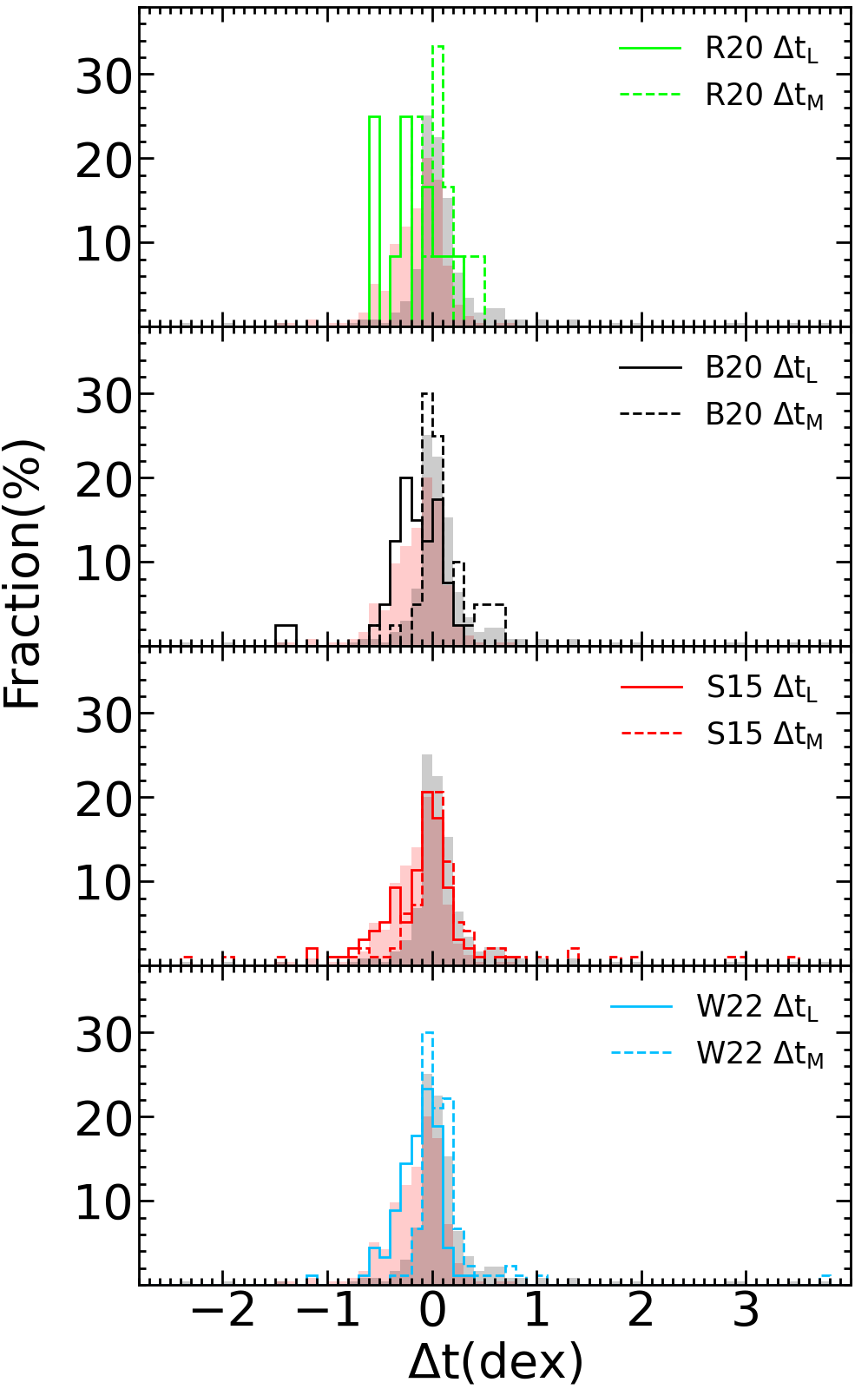}
    \caption{Fractional distributions of the difference between the light-weighted ($\Delta t_L$; solid line) and mass-weighted ($\Delta t_M$; dash line) ages from the the two fitting models (with/without a PL component). Line colors denote different samples same as Fig. \ref{fig1}, filled histograms represent distributions of whole samples in light-weighted (pink) and mass-weighted (gray).
\label{fig7}}
\end{figure}

As shown in \S\ref{sec:bpt}, the optical emission lines may be dominated by the host galaxy for our sample objects, and thus the AGN should only have a minor contribution to the spectral light. Considering this fact, we also carried out SSP synthesis without an AGN component (i.e., the pure stellar population model). In addition, this process can help us to estimate to what extent an FC component would affect our synthesis results. In Table \ref{tab3} we list the derived properties of the SSPs for different (sub)samples.

\begin{figure}
\centering
    \includegraphics[width=0.47\textwidth, angle=0]{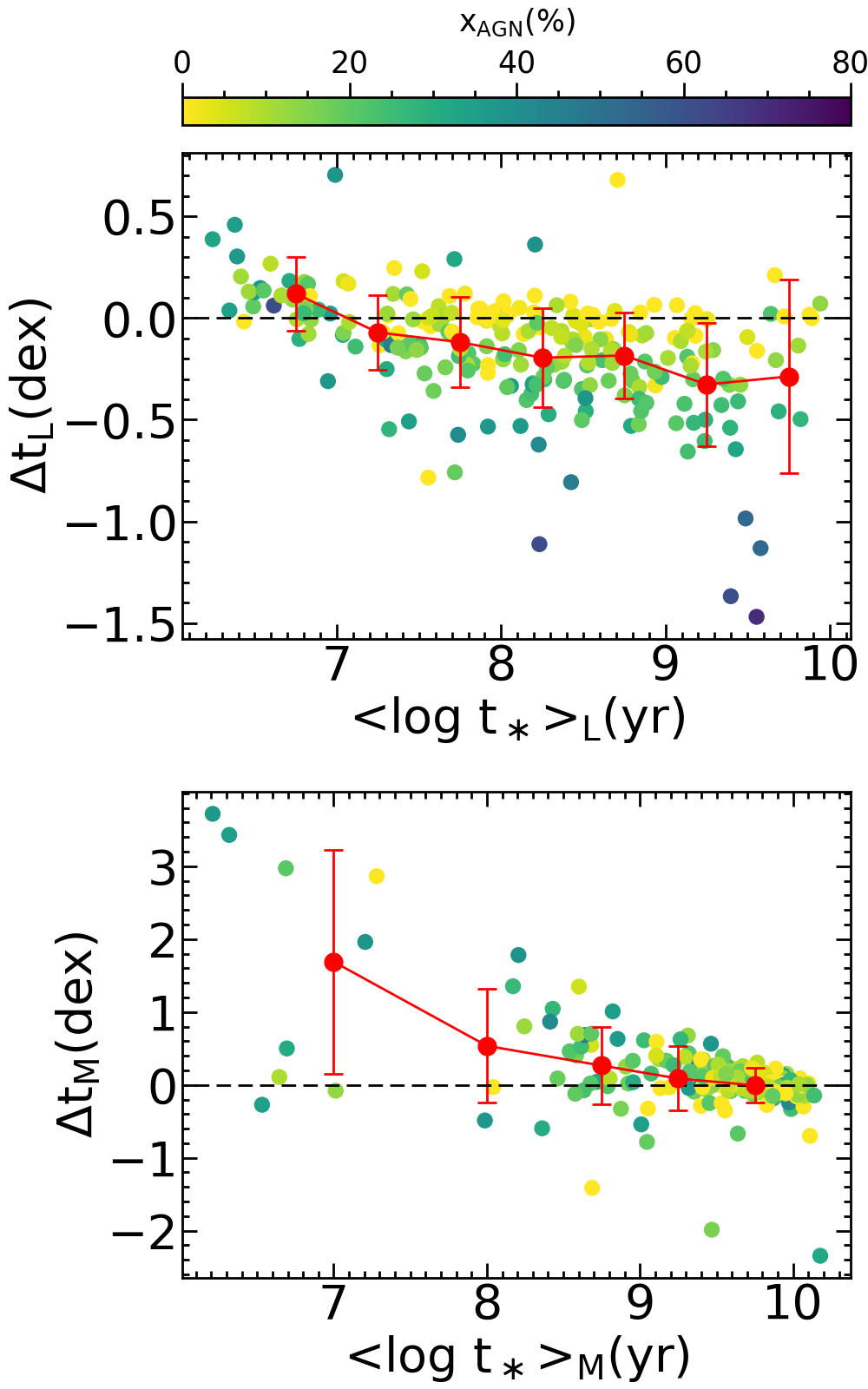}
    \caption{Difference between the mean stellar ages from the two fitting models as a function of $\langle \log t_{\star}\rangle_{L}$ (top) and $\langle \log t_{\star}\rangle_{M}$) (bottom). Samples colored by $x_{\rm {AGN}}$, and red solid circles with error bars show the median value and dispersion of each $\langle \log t_{\star}\rangle$ bin.
\label{fig8}}
\end{figure}

Compared the values listed in Table \ref{tab2} with those in Table \ref{tab3}, we find that, in general, the mass (light)-weighted age from the pure stellar population model would be older (younger) by $\sim$0.1 ($\sim$0.2) dex, and the mass (light)-weighted stellar metallicity would be reduced by a factor of 10$-$20\% (10$-$25\%). These results can be seen more clearly in Fig. \ref{fig7}, which plots the fractional distribution of $\Delta t=\langle \log t_{\star}\rangle_{\mathrm {noAGN}}-\langle \log t_{\star}\rangle_{\mathrm {AGN}}$, the difference of stellar ages between the pure stellar population and AGN-included models, for different samples. Fig. \ref{fig7} demonstrates that only a small fraction of galaxies have $\Delta t > 0.5$ dex, and each sample has a comparable dispersion in $\Delta t$. For the whole sample, both $\Delta t_M$ (filled gray histogram) and $\Delta t_L$ (filled pink histogram) peak around 0.0, and a Gaussian function can generally fit the distribution (except for the small fraction of galaxies showing large  $\Delta t$), which gives $(\mu,\sigma)_{\Delta t_M}=(0.02,0.13)~\mathrm{dex}$, and $(\mu,\sigma)_{\Delta t_L}=(0.10,0.19)~\mathrm{dex}$. 

We also note that for a few sources $\Delta t_M$ ($\Delta t_L$) can be as large as 3.8 (1.5) dex. These galaxies might have a large contribution from the AGN. To further check the AGN effect on our results, we plot $\Delta t$ as a function of the corresponding stellar age obtained from the AGN-included model, colorcoded by $x_{\rm {AGN}}$, as shown in Fig. \ref{fig8}. Indeed, 16 out of 24 sources with $\Delta t_M>0.5$ dex have $x_{\rm {AGN}}>20\%$, while a larger fraction (21/26) of sources with $|\Delta t_L|>0.5$ dex show $x_{\rm {AGN}}>20\%$. Furthermore, 9 out of 13 galaxies with $\Delta t_M<-0.3$ dex have $x_{\rm {AGN}}>15\%$, which is unexpected theoretically. Combining this with the fact that there also exist 39 sources (the ``S3" sample in Tables \ref{tab2} and \ref{tab3}) having $x_{\rm {AGN}}=0$, the dusty starburst-FC degeneracy, which would mistake a dusty starburst as an AGN FC and vice versa (\citealt{Cid2004}), has affected the results for about 20\% of our sample galaxies.

\citet{Cardoso2017} constructed synthetic galaxy spectra representing a wide range of galactic SFHs and including different PL contributions from AGNs, and performed a quantitative examination of the impact of the featureless continuum of an AGN on the estimation of physical properties of galaxies. The authors show that, at the empirical AGN detection threshold of $x_{\mathrm{AGN}} \simeq0.26$, the exclusion of a PL component from spectral fitting can lead to an overestimation by up to $\sim$4 dex in $\langle \log t_{\star}\rangle_M$ for the youngest ($t_M \lesssim 10^6$ yr) objects, and to an overestimation (underestimation) by up to $\sim$0.8 dex in $\langle \log t_{\star}\rangle_L$ for $t_L \lesssim 10^7$ yr ($t_L > 10^7$ yr). Furthermore, these biases become more severe with increasing $x_{\mathrm{AGN}}$ and are independent of the adopted SFH. As demonstrated in Fig. \ref{fig8}, our results agree very well with the findings in \citet{Cardoso2017}. We can see that $\Delta t$ is related to the corresponding mean stellar age, and $\Delta t_L$ (top panel) is more sensitive to $x_{\rm {AGN}}$ than $\Delta t_M$ (bottom panel). In particular, the transition of $\Delta t_L$ for our sample galaxies occurs at $\langle \log t_{\star}\rangle_{L} \sim 7$, same as that in \citet{Cardoso2017}. Therefore, the uncertainties in the mean stellar ages are generally in the range of $0.2-0.3$ dex, and there should be a tiny/mild effect on the stellar population results for the vast majority of our sample galaxies if we exclude the PL component from the model library, since $x_{\mathrm{AGN}}$ has a mean value of 0.16 for our whole sample. However, the large difference between these two models for some sources emphasizes the importance of the PL component in stellar population synthesis for AGN-hosts, and it is required a careful evaluation when the host galaxy spectrum is severely affected by an AGN.

\subsection{Connection Between AGN Activities and SFHs} \label{sec:Connection Between AGN Activities and SFHs}

\begin{figure}[tb]
\centering
    \includegraphics[width=0.47\textwidth, angle=0]{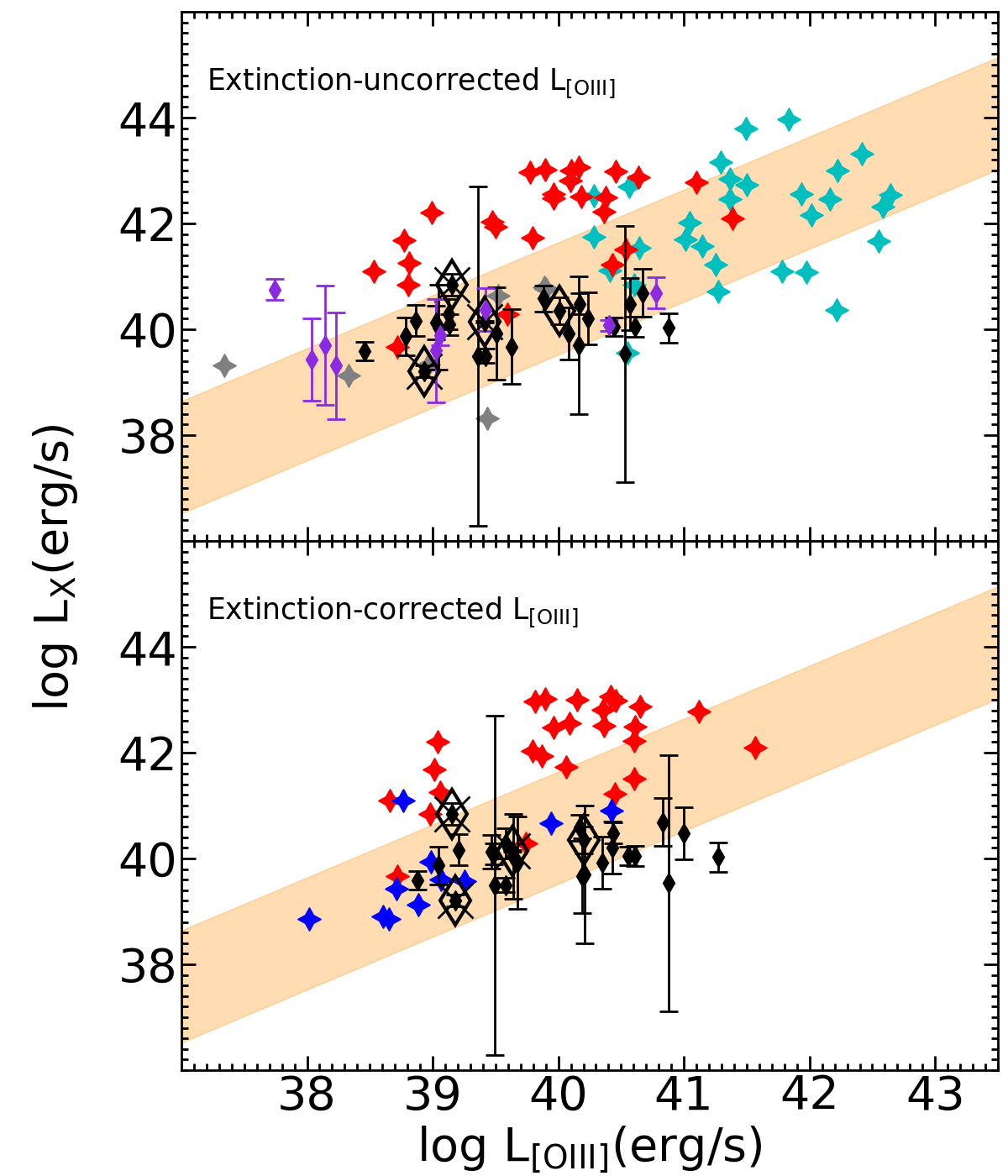}
    \caption{The observed X-ray luminosity $L_{\rm X}$ plotted against the extinction-uncorrected (top panel) and extinction-corrected (bottom panel) \OIII\ luminosity \LOIII. Diamonds represent the 35 sources in the X-ray ($L_{\mathrm{X}}=L_{2-12\,\mathrm{keV}}$) sample, with the black (violet) points included (excluded) in Fig. \ref{fig2}. Colored stars are non-Compton-thick type-2 AGNs (\citealp[cyan:][]{Heckman2005}; \citealp[red:][]{Gu2006}; and \citealp[blue:][]{Panessa2006}) and optical variability-selected low-mass AGNs (\citealp[gray;][]{Messick2023}) with $L_{\mathrm{X}}=L_{2-10\,\mathrm{keV}}$. The color-filled region shows the mean $L_{\mathrm{X}}/L_{\rm [O\,{\scriptsize \textsc{iii}}]}$ with $\pm 1 \sigma$ dispersion from \citet{Heckman2005}. Crosses and larger symbols are the same as Fig. \ref{fig2}
\label{fig9}}
\end{figure}

As one of the strongest optical lines in most galaxies, the \OIIII\ emission is sensitive to hardness and intensity of the radiation field, and thus its luminosity (\LOIII) can be used as a good tracer of the strength of AGN activities \citep{Kauffmann2003}. For the optical sample, C20 find a mild (anti-)correlation between $x_{\rm Y}$ ($\langle \log t_{\star}\rangle_{L}$) and \LOIII, especially when \LOIII$>10^{39}$~erg~s$^{-1}$, suggesting some physical connections between the nuclear star formation and AGN activities. Since the current sample is selected with multiple methods, and is about 1.5 times larger than that used in C20, we perform a similar analysis to C20 in the following, to further explore whether such correlation still holds for a more representative sample of AGN-host dwarfs.

To obtain an extinction-free luminosity of the \OIII\ line, we first estimate the intrinsic nebulae extinction, $A_{V,\,{\rm neb}}$, by utilizing the Balmer decrement as in \citet{Zhao2011} and C20. For the 9 Seyfer/Composite sources (see Fig. \ref{fig2}), we adopt an intrinsic ratio of $I_{\mathrm{H}\alpha}/I_{\mathrm{H}\beta}=3.1$, appropriate for AGNs \citep[e.g.,][]{Osterbrock2006,Gaskell1984}, for an electron temperature of 10$^4$~K and an electron density of 100~cm$^{-3}$. For the remaining sources located in the star-forming region in the BPT diagram, we take $I_{\mathrm{H}\alpha}/I_{\mathrm{H}\beta}=2.86$, suitable for \HII\ galaxies \citep{Brocklehurst1971}. For those objects with the observed Balmer decrement smaller than the intrinsic value, we assign their $A_{V,\,{\rm neb}}=0$. Then we correct the observed \OIII\ flux using the obtained $A_{V,\,{\rm neb}}$ and the \citet{Cardelli1989} reddening law. We find that $\sim73\%$ of our sample galaxies are weak AGNs having \LOIII$ < 10^7~L_{\odot}$ \citep{Kauffmann2003}.

Most of our sources are located in the star-forming region, though the BPT diagram might not be usable in classifying dwarf galaxies as discussed in \S\ref{sec:bpt}, and thereby it is worth investigating the origin of the \OIII\ emission for our sample galaxies. To this end, we check whether the relation between the X-ray luminosity ($L_{\rm X}$) and \LOIII for our sample galaxies, as shown in Fig. \ref{fig9}, follows the trend established by more massive objects, such as type-2 Seyferts, in which the X-ray luminosity is believed to be dominated by the AGN. It can be seen from the figure that, our X-ray sample has a similar $L_{\mathrm{X}}/L_{\rm [O\,{\scriptsize \textsc{iii}}]}$ ratio to those AGNs in \citet{Heckman2005} and \citet{Panessa2006}. However, the overall $L_{\mathrm{X}}/L_{\rm [O\,{\scriptsize \textsc{iii}}]}$ ratio of the Seyfert 2 AGNs from \citet{Gu2006} is about an order of magnitude larger than the other samples. Though the X-ray sample shows nuclear X-ray activity indicative of an AGN, it seems difficult to rule out the possibility that the host galaxy may heavily contaminate the \OIII\ luminosity based only the observed $L_{\mathrm{X}}/L_{\rm [O\,{\scriptsize \textsc{iii}}]}$ ratios. However, we note that for star-forming blue compact dwarf galaxies, no correlation between \LOIII\ and $\langle \log t_{\star}\rangle_{L}$ (or $x_{\mathrm{Y}}$) is found in C20.

\begin{table}[tb]
\bc
\begin{minipage}[]{140mm}
\caption[]{Spearman Correlation Coefficients\label{tab4}}\end{minipage}
\setlength{\tabcolsep}{6pt}
\small
 \begin{tabular}{ccccccc}
  \hline\noalign{\smallskip}
Sample  &  Number & \multicolumn{2}{c}{\LOIII-$\langle \log t_{\star}\rangle_{L}$} & \multicolumn{2}{c}{\LOIII-$x_{\rm Y}$} \\
        & & $\rho$ & $p$-value  & $\rho$    & $p$-value \\
    \hline\noalign{\smallskip}
radio   &  10 & 0.30 & 4.0E-01 & 0.16 & 6.5E-01 \\
X-ray   &  26 & -0.49 & 1.1E-02 & 0.46 & 1.8E-02 \\
mid-IR  &  79 & -0.50 & 2.9E-06 & 0.36 & 1.3E-03 \\
variability&  70 & -0.18 & 1.3E-01 & 0.10 & 3.9E-01 \\
All     &  181 & -0.45 & 3.4E-10 & 0.37 & 2.4E-07 \\
         \LOIII$>10^{39}$ & 139 & -0.47 & 6.7E-09 & 0.41 & 3.8E-07 \\
optical  & 136 & -0.20 & 1.8E-02 & 0.19 & 2.6E-02 \\
         \LOIII$>10^{39}$ & 91 & -0.41 & 6.4E-05 & 0.42 & 2.8E-05 \\
  \noalign{\smallskip}\hline
\end{tabular}
\ec
\end{table}

\begin{figure}[tb]
\centering
    \includegraphics[width=0.47\textwidth, angle=0]{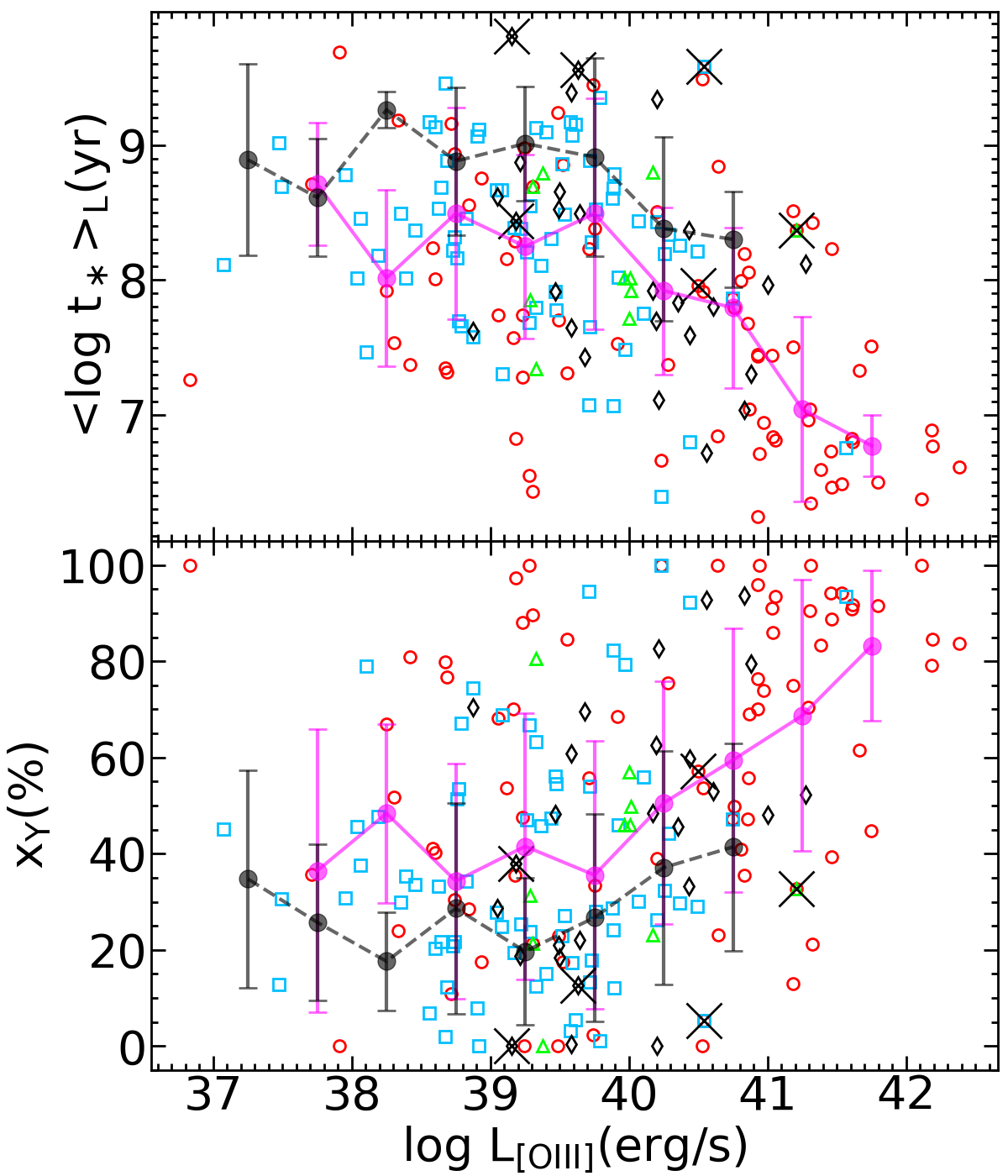}
    \caption{Extinction-corrected \LOIII\ vs. $\langle \log t_{\star}\rangle_{L}$ (top panel) and $x_{\rm Y}$ (bottom panel). Magenta solid and black dashed lines show the median (mean for $x_{\rm Y}$) trends for our and the optical samples, respectively. Colors and symbols are the same as Fig. \ref{fig2}.
\label{fig10}}
\end{figure}

Bearing the above caveat in mind, we plot the extinction-corrected \LOIII\ against the light-weight stellar age (upper panel) and fractional contribution of the young population (bottom panel) in Fig. \ref{fig10}, and list the Spearman correlation coefficients ($\rho$) and $p$-values in Table \ref{tab4}, for various (including the optical) samples. We can see that the current sample extends to a much (about one order of magnitude) higher  \LOIII\ compared with the opitcal sample studied in C20, but they show a similar trend (with an offset in $y$-axis). As indicated by the Spearman correlation coefficients, there exists a mild (anti-)correlation between \LOIII\ and $x_{\rm Y}$ ($\langle \log t_{\star}\rangle_{L}$) for the X-ray, mid-IR, $L_{\rm [O\,{\scriptsize \textsc{iii}}]}>10^{39}$~erg~s$^{-1}$ and whole samples, further suggesting a connection between AGN and star-forming activities.

Meanwhile, no correlation is found for the radio sample, which should be due to the fact there is a large scatter in the relation and the sample size is very small. Like the optical one, there is no correlation for the variability sample too, because it generally has the lowest \LOIII\ and most of the sample galaxies locate within the flat part of the trend, as illustrated in Fig. \ref{fig10}. To reach a solid conclusion regarding the correlation between AGN activity and SFH, a large sample selected with various methods and IFU data (such as MaNGA) is needed to reduce the influence of the host galaxy on the \OIII\ luminosity.

\subsection{Extinction} \label{Extinction}

For the stellar visual extinction $A_{V,\,\star}$, it is returned by STARLIGHT, modeled as due to a foreground dust screen using the continuum spectrum. To estimate the internal extinction suffered by nebular lines, we adopt equation (5) in \cite{Zhao2011}, namely
\[
    A_{V,\mathrm{neb}} = 7.93\times \log \left ( \frac{F_{\mathrm{H}\alpha}/F_{\mathrm{H}\beta}}{I_{\mathrm{H}\alpha}/I_{\mathrm{H}\beta}} \right ),
\]
where $F_{\rm H\alpha}/F_{\rm H\beta}$ and $I_{\rm H\alpha}/I_{\rm H\beta}$ are the observed and intrinsic Balmer decrements, respectively. Here we have assumed that the intrinsic Balmer line ratios are equal to Case B recombination, and used the \cite{Calzetti1994} reddening law. As mentioned in Section \ref{sec:Connection Between AGN Activities and SFHs}, two different $I_{{\rm H}\alpha}/I_{{\rm H}\beta}$ ratios are adopted respectively for sources located in the AGN/composite ($I_{{\rm H}\alpha}/I_{{\rm H}\beta}=3.1$) and SF ($I_{{\rm H}\alpha}/I_{{\rm H}\beta}=2.86$) regions in the BPT diagram.

To measure $A_{V,\,\mathrm{neb}}$ of a galaxy, we require that both H$\alpha$ and H$\beta$ have $\mathrm{S/N}>3$. For the radio, X-ray, mid-IR, variability and whole sample, the median values of $\log(F_{\rm H\alpha}/F_{\rm H\beta})$ are 0.55$\pm$0.07, 0.54$\pm$0.05, 0.51$\pm$0.05, 0.56$\pm$0.06 and 0.54$\pm$0.07, respectively, corresponding to the median $A_{V,\,\mathrm{neb}}$ of 0.74$\pm$0.57, 0.59$\pm$0.32, 0.45$\pm$0.36, 0.90$\pm$0.43 and 0.66$\pm$0.48 mag. Except for variability sample, other samples show smaller nebular extinction than the optical sample ($A_{V,\,{\rm neb}}=0.9$; \citealp{Cai2020}), but are comparable to the star-forming galaxies with similar stellar masses presented in \cite{2010MNRAS.409..421G}.

\citet{Calzetti1994} find that the extinction derived from the Balmer emission line is about twice as much as that from the continuum \citep[see also][]{Charlot2000}. For our sources with positive $A_{V,\,{\rm neb}}$ and $A_{V,\,\star}$, the median values of $A_{V,\,{\rm neb}} / A_{V,\,\star}$ are 1.60$\pm$1.15 (Radio), 1.74$\pm$1.14 (X-ray), 1.59$\pm$1.42 (mid-IR), 1.84$\pm$1.02 ( Variability), and 1.72$\pm$1.25 (whole sample), which are 10\%-20\% smaller than the value found for nearby star-forming galaxies (\citealp{Calzetti2000}). Our results seem consistent with the finding in \cite{Woo2024}, who study the model uncertainties of STARLIGHT and show that the stellar extinction tends to be overestimated when $\mathrm{S/N} \lesssim15$ (see their Fig. 12). At $\mathrm{S/N}=10$, $\Delta E(B-V)$ is about 0.02 dex (\citealp[]{Woo2024}), resulting in an overestimation of $A_{V,\,\star}\sim 5\%$, which is a bit smaller than our value of 10-20\%.

Therefore, we further check other possible reason causing the above discrepancy, and find that there exists an obvious difference in $A_{V,\,{\rm neb}} / A_{V,\,\star}$ between the two subsamples S3 ($x_{\rm {AGN}}=0$; 22 sources) and S4 ($x_{\rm {AGN}}>0$; 139 sources), with the median $A_{V,\,{\rm neb}} / A_{V,\,\star}$ ratios of 1.35 and 1.76, respectively. This result implies that the stellar extinction for the S3 subsample could be overestimated to some degree, which can be naturally explained by the degeneracy between AGN FC and dusty starburst, in the sense that STARLIGHT would have treated the AGN FC as a dusty starburst. To check to what extent the age-extinction and/or AGN FC-dusty starburst degeneracy would affect our fitting results, we re-run STARLIGHT using the same parameters except limiting $A_{V,\,\star}$ to the value of $A_{V,\,{\rm neb}}$ for 23 sources with $A_{V,\,\star}/A_{V,\,{\rm neb}}>1.5$. We find that the mean stellar age increases by a mean value of $\Delta \langle \log t_{\star}\rangle_{L}=0.30\pm0.18$ ($\Delta \langle \log t_{\star}\rangle_{M}=0.45\pm0.72$) dex.

\section{Summary} \label{sec:Summary}

In this paper, we present detailed results of our stellar population synthesis for a sample of 235 AGN-host dwarf galaxies selected by different methods from radio, X-ray, mid-IR and variability using the STARLIGHT code and fiber spectra from SDSS DR8. Our main results can be summarized as follows:

1. Among our four samples, the mid-IR (variability) sample is the youngest (oldest) with the median value of the light-weighted mean stellar age of $10^{7.74}\ (10^{8.44})$ yr, and 57.7\% (23.3\%) of the sample sources have $\langle \log t_{\star}\rangle_{L}<8$. Our sample galaxies are $\sim$0.7 dex younger than the optically selected AGN-host dwarfs, but they have similar distributions in $\sigma_{L}(\log t_{\star})$, suggesting complex and diverse SFHs in all of the AGN-host dwarf systems, irrespective of the selection method.

2. AGNs are unlikely to strongly influence the chemical evolution of their host dwarf galaxies, as evidenced by that they show a similar metallicity-mass relation and a flat radial profile of $\langle Z_{\star}\rangle_{L}$ within 1$R_e$ to normal dwarfs.

3. For our whole sample, there is a negative gradient of $\langle \log t_{\star}\rangle_{L}$, indicative of inside-out and/or suppression of star-formation in the central region. Only $\sim$9.5\% of the sources with stellar mass of $10^9-3\times
10^9~M_\odot$ have $\mu_{\mathrm{Y}}>5\%$, far less than the fraction of 25$-$40\% in literature for normal dwarf galaxies, suggesting that AGNs in dwarfs seem to suppress the current star formation of the host galaxies.

4. We compare the difference between the pure stellar population model and AGN-included model, and find that our results agree very well with the previous findings in literature. The mass (light)-weighted age from the pure stellar population model would be older (younger) by $\sim$0.1 ($\sim$0.2) dex, and the mass (light)-weighted stellar metallicity would be reduced by a factor of 10$-$20\% (10$-$25\%). 

5. Like the optical sample, there exists a mild (anti-)correlation between \LOIII\ and $x_{\rm Y}$ ($\langle \log t_{\star}\rangle_{L}$) for the X-ray, mid-IR, $L_{\rm [O\,{\scriptsize \textsc{iii}}]}>10^{39}$~erg~s$^{-1}$ and whole samples, further suggesting there are some connections between the AGN and star-forming activities.

\normalem
\begin{acknowledgements}
We thank the anonymous refree for careful reading and thoughtful comments which improved the paper. The work is supported by the China Manned Space Project with No. CMS-CSST-2021-A06, the National Key R\&D Program of China with No. 2021YFA1600404, the Natural Science Foundation of China (NSFC; grant Nos. 12173079 and 11991051). The STARLIGHT project is supported by the Brazilian agencies CNPq, CAPES, and FAPESP and by the France–Brazil CAPES/ Cofecub program. All the authors acknowledge the work of the Sloan Digital Sky Survey (SDSS) team. Funding for SDSS-III has been provided by the Alfred P. Sloan Foundation, the Participating Institutions, the National Science Foundation, and the U.S. Department of Energy Office of Science. The SDSS-III website is http://www.sdss3.org/. SDSS-III is managed by the Astrophysical Research Consortium for the Participating Institutions of the SDSS-III Collaboration including the University of Arizona, the Brazilian Participation Group, Brookhaven National Laboratory, Carnegie Mellon University, University of Florida, the French Participation Group, the German Participation Group, Harvard University, the Instituto de Astrofisica de Canarias, the Michigan State/Notre Dame/JINA Participation Group, Johns Hopkins University, Lawrence Berkeley National Laboratory, Max Planck Institute for Astrophysics, Max Planck Institute for Extraterrestrial Physics, New Mexico State University, New York University, Ohio State University, Pennsylvania State University, University of Portsmouth, Princeton University, the Spanish Participation Group, University of Tokyo, University of Utah, Vanderbilt University, University of Virginia, University of Washington, and Yale University.
\end{acknowledgements}
  
\bibliographystyle{raa}
\bibliography{bibtex}

\begin{thebibliography}{97}
\providecommand\natexlab[1]{#1}
\providecommand\JournalTitle[1]{#1}

\bibitem[Aihara {et~al.}(2011)]{Aihara2011}
Aihara, H., Allende~Prieto, C., An, D., {et~al.} 2011, \apjs, 193, 29

\bibitem[Baldwin {et~al.}(1981)]{Baldwin1981}
Baldwin, J.~A., Phillips, M.~M., \& Terlevich, R. 1981, \pasp, 93, 5

\bibitem[Battaglia {et~al.}(2006)]{Battaglia2006}
Battaglia, G., Tolstoy, E., Helmi, A., {et~al.} 2006, \aap, 459, 423

\bibitem[Belfiore {et~al.}(2018)]{Belfiore2018}
Belfiore, F., Maiolino, R., Bundy, K., {et~al.} 2018, \mnras, 477, 3014

\bibitem[Bellovary {et~al.}(2011)]{Bellovary2011}
Bellovary, J., Volonteri, M., Governato, F., {et~al.} 2011, \apj, 742, 13

\bibitem[Bernard {et~al.}(2008)]{Bernard2008}
Bernard, E.~J., Gallart, C., Monelli, M., {et~al.} 2008, \apj, 678, L21

\bibitem[Birchall {et~al.}(2020)]{Birchall2020}
Birchall, K.~L., Watson, M.~G., \& Aird, J. 2020, \mnras, 492, 2268

\bibitem[Blanton {et~al.}(2011)]{Blanton2011}
Blanton, M.~R., Kazin, E., Muna, D., Weaver, B.~A., \& Price-Whelan, A. 2011, \aj, 142, 31

\bibitem[Brocklehurst(1971)]{Brocklehurst1971}
Brocklehurst, M. 1971, \mnras, 153, 471

\bibitem[Bruzual \& Charlot(2003)]{Bruzual2003}
Bruzual, G., \& Charlot, S. 2003, \mnras, 344, 1000

\bibitem[Cai {et~al.}(2021)]{Cai2021}
Cai, W., Zhao, Y.-H., \& Bai, J.-M. 2021, RAA, 21, 204

\bibitem[Cai {et~al.}(2020)]{Cai2020}
Cai, W., Zhao, Y., Zhang, H.-X., Bai, J.-M., \& Liu, H.-T. 2020, \apj, 903, 58

\bibitem[Calzetti {et~al.}(2000)]{Calzetti2000}
Calzetti, D., Armus, L., Bohlin, R.~C., {et~al.} 2000, \apj, 533, 682

\bibitem[Calzetti {et~al.}(1994)]{Calzetti1994}
Calzetti, D., Kinney, A.~L., \& Storchi-Bergmann, T. 1994, \apj, 429, 582

\bibitem[{Cann} {et~al.}(2019)]{2019ApJ...870L...2C}
{Cann}, J.~M., {Satyapal}, S., {Abel}, N.~P., {et~al.} 2019, \apjl, 870, L2

\bibitem[Cardelli {et~al.}(1989)]{Cardelli1989}
Cardelli, J.~A., Clayton, G.~C., \& Mathis, J.~S. 1989, \apj, 345, 245

\bibitem[Cardoso {et~al.}(2017)]{Cardoso2017}
Cardoso, L. S.~M., Gomes, J.~M., \& Papaderos, P. 2017, \aap, 604, A99

\bibitem[Charlot \& Fall(2000)]{Charlot2000}
Charlot, S., \& Fall, S.~M. 2000, \apj, 539, 718

\bibitem[Cid~Fernandes {et~al.}(2004)]{Cid2004}
Cid~Fernandes, R., Gu, Q., Melnick, J., {et~al.} 2004, \mnras, 355, 273

\bibitem[Cid~Fernandes {et~al.}(2005)]{Cid2005}
Cid~Fernandes, R., Mateus, A., Sodré, L., Stasińska, G., \& Gomes, J.~M. 2005, \mnras, 358, 363

\bibitem[Davies {et~al.}(2017)]{Davies2017}
Davies, B., Kudritzki, R.-P., Lardo, C., {et~al.} 2017, \apj, 847, 112

\bibitem[Dolphin {et~al.}(2005)]{Dolphin2005}
Dolphin, A.~E., Weisz, D.~R., Skillman, E.~D., \& Holtzman, J.~A. 2005, Star Formation Histories of Local Group Dwarf Galaxies

\bibitem[Dubois {et~al.}(2013)]{Dubois2013}
Dubois, Y., Gavazzi, R., Peirani, S., \& Silk, J. 2013, \mnras, 433, 3297

\bibitem[Dubois {et~al.}(2016)]{Dubois2016}
Dubois, Y., Peirani, S., Pichon, C., {et~al.} 2016, \mnras, 463, 3948

\bibitem[Gaibler {et~al.}(2012)]{Gaibler2012}
Gaibler, V., Khochfar, S., Krause, M., \& Silk, J. 2012, \mnras, 425, 438

\bibitem[Gallart {et~al.}(2015)]{Gallart2015}
Gallart, C., Monelli, M., Mayer, L., {et~al.} 2015, \apj, 811, L18

\bibitem[{Garn} \& {Best}(2010)]{2010MNRAS.409..421G}
{Garn}, T., \& {Best}, P.~N. 2010, \mnras, 409, 421

\bibitem[Gaskell \& Ferland(1984)]{Gaskell1984}
Gaskell, C.~M., \& Ferland, G.~J. 1984, \pasp, 96, 393

\bibitem[Ge {et~al.}(2018)]{Ge2018}
Ge, J., Yan, R., Cappellari, M., {et~al.} 2018, \mnras, 478, 2633

\bibitem[González~Delgado {et~al.}(2015)]{Gonzalez2015}
González~Delgado, R.~M., García-Benito, R., Pérez, E., {et~al.} 2015, \aap, 581, A103

\bibitem[Groves {et~al.}(2006)]{Groves2006}
Groves, B.~A., Heckman, T.~M., \& Kauffmann, G. 2006, \mnras, 371, 1559

\bibitem[Gu {et~al.}(2006)]{Gu2006}
Gu, Q., Melnick, J., Cid~Fernandes, R., {et~al.} 2006, \mnras, 366, 480

\bibitem[Harbeck {et~al.}(2001)]{Harbeck2001}
Harbeck, D., Grebel, E.~K., Holtzman, J., {et~al.} 2001, \aj, 122, 3092

\bibitem[Heckman {et~al.}(2005)]{Heckman2005}
Heckman, T.~M., Ptak, A., Hornschemeier, A., \& Kauffmann, G. 2005, \apj, 634, 161

\bibitem[Hodge(1989)]{Hodge1989}
Hodge, P. 1989, \araa, 27, 139

\bibitem[Ishibashi \& Fabian(2012)]{Ishibashi2012}
Ishibashi, W., \& Fabian, A.~C. 2012, \mnras, 427, 2998

\bibitem[Jarrett {et~al.}(2011)]{Jarrett2011}
Jarrett, T.~H., Cohen, M., Masci, F., {et~al.} 2011, \apj, 735, 112

\bibitem[{Jin} {et~al.}(2022)]{2022ApJ...926..184J}
{Jin}, J.-J., {Wu}, X.-B., \& {Feng}, X.-T. 2022, \apj, 926, 184

\bibitem[Jin {et~al.}(2022)]{Jinwu2022}
Jin, J.-J., Wu, X.-B., \& Feng, X.-T. 2022, \apj, 926, 184

\bibitem[{Jin} {et~al.}(2018)]{2018ApJ...864...32J}
{Jin}, J.-J., {Zhu}, Y.-N., {Meng}, X.-M., {Lei}, F.-J., \& {Wu}, H. 2018, \apj, 864, 32

\bibitem[Kalfountzou {et~al.}(2012)]{Kalfountzou2012}
Kalfountzou, E., Jarvis, M.~J., Bonfield, D.~G., \& Hardcastle, M.~J. 2012, \mnras, 427, 2401

\bibitem[Kalfountzou {et~al.}(2014)]{Kalfountzou2014}
Kalfountzou, E., Stevens, J.~A., Jarvis, M.~J., {et~al.} 2014, \mnras, 442, 1181

\bibitem[Kauffmann(2014)]{Kauffmann2014}
Kauffmann, G. 2014, \mnras, 441, 2717

\bibitem[Kauffmann {et~al.}(1993)]{Kauffmann1993}
Kauffmann, G., White, S. D.~M., \& Guiderdoni, B. 1993, \mnras, 264, 201

\bibitem[Kauffmann {et~al.}(2003)]{Kauffmann2003}
Kauffmann, G., Heckman, T.~M., Tremonti, C., {et~al.} 2003, \mnras, 346, 1055

\bibitem[Kewley {et~al.}(2013)]{Kewley2013}
Kewley, L.~J., Dopita, M.~A., Leitherer, C., {et~al.} 2013, \apj, 774, 100

\bibitem[Kewley {et~al.}(2001)]{Kewley2001a}
Kewley, L.~J., Dopita, M.~A., Sutherland, R.~S., Heisler, C.~A., \& Trevena, J. 2001, \apj, 556, 121

\bibitem[Kewley {et~al.}(2005)]{Kewley2005}
Kewley, L.~J., Jansen, R.~A., \& Geller, M.~J. 2005, \pasp, 117, 227

\bibitem[Kirby {et~al.}(2013)]{Kirby2013}
Kirby, E.~N., Cohen, J.~G., Guhathakurta, P., {et~al.} 2013, \apj, 779, 102

\bibitem[Lehmer {et~al.}(2016)]{Lehmer2016}
Lehmer, B.~D., Basu-Zych, A.~R., Mineo, S., {et~al.} 2016, \apj, 825, 7

\bibitem[Li {et~al.}(2015)]{Li2015}
Li, C., Wang, E., Lin, L., {et~al.} 2015, \apj, 804, 125

\bibitem[Ludwig {et~al.}(2012)]{Ludwig2012}
Ludwig, R.~R., Greene, J.~E., Barth, A.~J., \& Ho, L.~C. 2012, \apj, 756, 51

\bibitem[{Mahoro} {et~al.}(2022)]{Mahoro2022}
{Mahoro}, A., {Povi{\'c}}, M., {V{\"a}is{\"a}nen}, P., {Nkundabakura}, P., \& {van der Heyden}, K. 2022, \mnras, 513, 4494

\bibitem[Maiolino {et~al.}(2017)]{Maiolino2017}
Maiolino, R., Russell, H.~R., Fabian, A.~C., {et~al.} 2017, Nature, 544, 202

\bibitem[{Mallmann} {et~al.}(2018)]{2018MNRAS.478.5491M}
{Mallmann}, N.~D., {Riffel}, R., {Storchi-Bergmann}, T., {et~al.} 2018, \mnras, 478, 5491

\bibitem[Marzke \& da~Costa(1997)]{Marzke1997}
Marzke, R.~O., \& da~Costa, L.~N. 1997, \aj, 113, 185

\bibitem[Mateo(1998)]{Mateo1998}
Mateo, M.~L. 1998, \araa, 36, 435

\bibitem[Mateus {et~al.}(2006)]{Mateus2006}
Mateus, A., Sodré, L., Cid~Fernandes, R., {et~al.} 2006, \mnras, 370, 721

\bibitem[McQuinn {et~al.}(2011)]{McQuinn2011}
McQuinn, K. B.~W., Skillman, E.~D., Dalcanton, J.~J., {et~al.} 2011, \apj, 740, 48

\bibitem[McQuinn {et~al.}(2010)]{McQuinn2010}
McQuinn, K. B.~W., Skillman, E.~D., Cannon, J.~M., {et~al.} 2010, \apj, 721, 297

\bibitem[Messick {et~al.}(2023)]{Messick2023}
Messick, A., Baldassare, V., Geha, M., \& Greene, J. 2023, \apj, 953, 18

\bibitem[Mineo {et~al.}(2012)]{Mineo2012}
Mineo, S., Gilfanov, M., \& Sunyaev, R. 2012, \mnras, 426, 1870

\bibitem[Moran {et~al.}(2014)]{Moran2014}
Moran, E.~C., Shahinyan, K., Sugarman, H.~R., Vélez, D.~O., \& Eracleous, M. 2014, \aj, 148, 136

\bibitem[Morelli {et~al.}(2015)]{Morelli2015}
Morelli, L., Corsini, E.~M., Pizzella, A., {et~al.} 2015, \mnras, 452, 1128

\bibitem[Nayakshin \& Zubovas(2012)]{Nayakshin2012}
Nayakshin, S., \& Zubovas, K. 2012, \mnras, 427, 372

\bibitem[O'Donnell(1994)]{O'Donnell1994}
O'Donnell, J.~E. 1994, \apj, 422, 158

\bibitem[Oh {et~al.}(2011)]{Oh2011}
Oh, K., Sarzi, M., Schawinski, K., \& Yi, S.~K. 2011, \apjs, 195, 13

\bibitem[Osterbrock \& Ferland(2006)]{Osterbrock2006}
Osterbrock, D.~E., \& Ferland, G.~J. 2006, Astrophysics of gaseous nebulae and active galactic nuclei, 2nd (University Science Books, Sausalito, CA)

\bibitem[Panessa {et~al.}(2006)]{Panessa2006}
Panessa, F., Bassani, L., Cappi, M., {et~al.} 2006, \aap, 455, 173

\bibitem[Pillepich {et~al.}(2018)]{Pillepich2018}
Pillepich, A., Nelson, D., Hernquist, L., {et~al.} 2018, \mnras, 475, 648

\bibitem[Querejeta {et~al.}(2016)]{Querejeta2016}
Querejeta, M., Schinnerer, E., García-Burillo, S., {et~al.} 2016, \aap, 593, A118

\bibitem[Reines {et~al.}(2020)]{Reines2020}
Reines, A.~E., Condon, J.~J., Darling, J., \& Greene, J.~E. 2020, \apj, 888, 36

\bibitem[Reines {et~al.}(2013)]{Reines2013}
Reines, A.~E., Greene, J.~E., \& Geha, M. 2013, \apj, 775, 116

\bibitem[Roig {et~al.}(2015)]{Roig2015}
Roig, B., Blanton, M.~R., \& Yan, R. 2015, \apj, 808, 26

\bibitem[Rosen {et~al.}(2016)]{Rosen2016}
Rosen, S.~R., Webb, N.~A., Watson, M.~G., {et~al.} 2016, \aap, 590, A1

\bibitem[Sartori {et~al.}(2015)]{Sartori2015}
Sartori, L.~F., Schawinski, K., Treister, E., {et~al.} 2015, \mnras, 454, 3722

\bibitem[Schaye {et~al.}(2015)]{Schaye2015}
Schaye, J., Crain, R.~A., Bower, R.~G., {et~al.} 2015, \mnras, 446, 521

\bibitem[Schlegel {et~al.}(1998)]{Schlegel1998}
Schlegel, D.~J., Finkbeiner, D.~P., \& Davis, M. 1998, \apj, 500, 525

\bibitem[{Schutte} \& {Reines}(2022)]{2022Natur.601..329S}
{Schutte}, Z., \& {Reines}, A.~E. 2022, \nat, 601, 329

\bibitem[Silk(2013)]{Silk2013}
Silk, J. 2013, \apj, 772, 112

\bibitem[Silk \& Norman(2009)]{Silk2009}
Silk, J., \& Norman, C. 2009, \apj, 700, 262

\bibitem[Stern {et~al.}(2012)]{Stern2012}
Stern, D., Assef, R.~J., Benford, D.~J., {et~al.} 2012, \apj, 753, 30

\bibitem[Taibi {et~al.}(2022)]{Taibi2022}
Taibi, S., Battaglia, G., Leaman, R., {et~al.} 2022, \aap, 665, A92

\bibitem[Tolstoy {et~al.}(2009)]{Tolstoy2009}
Tolstoy, E., Hill, V., \& Tosi, M. 2009, \araa, 47, 371

\bibitem[Tolstoy {et~al.}(2004)]{Tolstoy2004}
Tolstoy, E., Irwin, M.~J., Helmi, A., {et~al.} 2004, \apj, 617, L119

\bibitem[Trump {et~al.}(2015)]{Trump2015}
Trump, J.~R., Sun, M., Zeimann, G.~R., {et~al.} 2015, \apj, 811, 26

\bibitem[Veilleux \& Osterbrock(1987)]{Veilleux1987}
Veilleux, S., \& Osterbrock, D.~E. 1987, \apjs, 63, 295

\bibitem[Vogelsberger {et~al.}(2014)]{Vogelsberger2014}
Vogelsberger, M., Genel, S., Springel, V., {et~al.} 2014, \mnras, 444, 1518

\bibitem[Ward {et~al.}(2022)]{Ward2022}
Ward, C., Gezari, S., Nugent, P., {et~al.} 2022, \apj, 936, 104

\bibitem[Weisz {et~al.}(2011{\natexlab{a}})]{Weisz2011b}
Weisz, D.~R., Dalcanton, J.~J., Williams, B.~F., {et~al.} 2011{\natexlab{a}}, \apj, 739, 5

\bibitem[Weisz {et~al.}(2011{\natexlab{b}})]{weisz2011a}
Weisz, D.~R., Dolphin, A.~E., Dalcanton, J.~J., {et~al.} 2011{\natexlab{b}}, \apj, 743, 8

\bibitem[{Woo} {et~al.}(2024)]{Woo2024}
{Woo}, J., {Walters}, D., {Archinuk}, F., {et~al.} 2024, arXiv e-prints, arXiv:2401.12300

\bibitem[Yan(2011)]{Yan2011}
Yan, R. 2011, \aj, 142, 153

\bibitem[Zahid {et~al.}(2017)]{Zahid2017}
Zahid, H.~J., Kudritzki, R.-P., Conroy, C., Andrews, B., \& Ho, I.~T. 2017, \apj, 847, 18

\bibitem[{Zhang} {et~al.}(2017)]{2017ApJS..233...13Z}
{Zhang}, H.-X., {Puzia}, T.~H., \& {Weisz}, D.~R. 2017, \apjs, 233, 13

\bibitem[Zhao {et~al.}(2011)]{Zhao2011}
Zhao, Y., Gu, Q., \& Gao, Y. 2011, \aj, 141, 68

\bibitem[Zheng {et~al.}(2017)]{Zheng2017}
Zheng, Z., Wang, H., Ge, J., {et~al.} 2017, \mnras, 465, 4572

\end{thebibliography}

\end{document}